\RequirePackage[2020-02-02]{latexrelease}
\documentclass[conference]{IEEEtran}
\IEEEoverridecommandlockouts
\usepackage{cite}
\usepackage{amsmath,amssymb,amsfonts}
\usepackage{algorithmic}
\usepackage{graphicx}
\usepackage{textcomp}
\usepackage{xcolor}
\usepackage{booktabs} 
\usepackage{subcaption}
\usepackage{listings}
\usepackage{tabularx} % in the preamble
\usepackage{url}
\usepackage[autostyle]{csquotes}  

\usepackage{titlesec}
\usepackage[printwatermark]{xwatermark}
\usepackage{graphicx}
\graphicspath{ {./images/} }
\def\BibTeX{{\rm B\kern-.05em{\sc i\kern-.025em b}\kern-.08em
    T\kern-.1667em\lower.7ex\hbox{E}\kern-.125emX}}

\usepackage{framed}
\usepackage{float}
\usepackage{lineno,hyperref}
\modulolinenumbers[5]

\begin{document}

\title{ECP SOLLVE: Validation and Verification Testsuite Status Update and 
Compiler Insight for OpenMP\\
\thanks{
This manuscript has been authored by UT-Battelle, LLC under Contract No. DE-AC05-00OR22725 with the U.S. Department of Energy.  The publisher, by accepting the article for publication, acknowledges that the U.S. Government retains a non-exclusive, paid up, irrevocable, world-wide license to publish or reproduce the published form of the manuscript, or allow others to do so, for U.S. Government purposes. The DOE will provide public access to these results in accordance with the DOE Public Access Plan (http://energy.gov/downloads/doe-public-access-plan).}
}

\author{\IEEEauthorblockN{1\textsuperscript{st} Thomas Huber}
\IEEEauthorblockA{\textit{University of Delaware} \\
thuber@udel.edu}
\and
\IEEEauthorblockN{2\textsuperscript{nd} Swaroop Pophale}
\IEEEauthorblockA{\textit{Oak Ridge National Laboratory} \\
pophaless@ornl.gov}
\and
\IEEEauthorblockN{3\textsuperscript{rd} Nolan Baker}
\IEEEauthorblockA{\textit{University of Delaware} \\
nolanb@udel.edu}
\and
\IEEEauthorblockN{4\textsuperscript{th} Michael Carr}
\IEEEauthorblockA{\textit{University of Delaware} \\
mjcarr@udel.edu}
\and
\IEEEauthorblockN{5\textsuperscript{th} Nikhil Rao}
\IEEEauthorblockA{\textit{University of Delaware} \\
nikhilr@udel.edu}
\and
\IEEEauthorblockN{6\textsuperscript{th} Jaydon Reap}
\IEEEauthorblockA{\textit{University of Delaware} \\
jreap@udel.edu}
\and
\IEEEauthorblockN{7\textsuperscript{th} Kristina Holsapple}
\IEEEauthorblockA{\textit{University of Delaware} \\
kris@udel.edu}
\and
\IEEEauthorblockN{8\textsuperscript{th} Joshua Hoke Davis}
\IEEEauthorblockA{\textit{University of Maryland} \\
jhdavis@umd.edu}
\and
\IEEEauthorblockN{9\textsuperscript{th} Tobias Burnus}
\IEEEauthorblockA{\textit{Siemens} \\
Tobias\_Burnus@mentor.com}
\and
\IEEEauthorblockN{10\textsuperscript{th} Seyong Lee}
\IEEEauthorblockA{\textit{Oak Ridge National Laboratory} \\
lees2@ornl.gov}
\and
\IEEEauthorblockN{11\textsuperscript{th} David E. Bernholdt}
\IEEEauthorblockA{\textit{Oak Ridge National Laboratory} \\
bernholdtde@ornl.gov}
\and
\IEEEauthorblockN{12\textsuperscript{th} Sunita Chandrasekaran}
\IEEEauthorblockA{\textit{University of Delaware} \\
schandra@udel.edu}
}

\maketitle
\thispagestyle{plain}
\pagestyle{plain}
\begin{abstract}
The OpenMP language continues to evolve with every new specification release, as does the need to validate and verify the new features that have been implemented by the different vendors. With the release of OpenMP 5.0 and OpenMP 5.1, plenty of new target offload and host-based features have been introduced to the programming model. While OpenMP continues to grow in maturity, there is an observable growth in the number of compiler and hardware vendors that support OpenMP.
In this manuscript, we focus on evaluating the conformity and implementation progress of various compiler vendors such as Cray, IBM, GNU, Clang/LLVM, NVIDIA, and Intel. We specifically address the 4.5, 5.0, and 5.1 versions of the speci-fication. For our experimental setup, we use the Crusher and Summit computing systems hosted by Oak Ridge National Lab’s Computing Facilities. The effort of fault-finding in these implementations is especially valuable for application developers who are using new OpenMP features to accelerate their scientific codes. We present insights into the current implementation status of various vendors, the progression of specific compiler’s support for OpenMP over-time, and examples of how our test suite has influenced discussion regarding the correct interpretation of the OpenMP specification. By evaluating OpenMP conformity of pre-Exascale computing systems, we aim to detail progress and status of AMD + Cray ecosystem before the system and their OpenMP implementation is used for mission critical applications when the first Exascale Computer Frontier is made available to researchers and scientists.
%researchers and scientists.
%SP: making the language softer to avoid antagonizing vendors
\end{abstract}

\begin{IEEEkeywords}
OpenMP, GPU, Offloading, LLVM
\end{IEEEkeywords}

\section{Introduction}
Seven out of the ten fastest supercomputers in the world are heterogeneous systems ~\cite{top500june2022}. Heterogeneous systems may be comprised of a CPU and an accelerator such as GPUs, FPGAs, APUs, etc., however, the top performing supercomputers tend to opt towards a configuration of CPU and GPU. For two years in a row (June 2020 - June 2022), the Fugaku A64FX supercomputer produced by Fujitsu and ARM and hosted by RIKEN Center for Computational Science held the title for the fastest supercomputer \cite{FugakuSystem} and proved that a CPU only configuration was able to transcend the performance of heterogeneous systems like Oak Ridge National Laboratory (ORNL)'s Summit (IBM Power9 CPU + NVIDIA V100 GPU). 

Following the release of ORNL's Frontier, the world's first Exascale supercomputer (HPL score of 1.102 Exaflop/s using 8,730,112 cores)~\cite{InsideHPC}, again the top supercomputer in the world is composed of a heterogeneous mix of compute power (3rd Gen AMD EPYC 64C CPUs and AMD Instinct MI250X GPU accelerators). 
As hardware vendors with heterogeneous systems in the TOP500, HPE Cray, IBM, Intel, NVIDIA, and AMD provide software support for various parallel programming models that allow users to port their parallel applications to accelerators. 

Considering the various changes in hardware architecture offerings over the years, programming models and base languages are incorporating parallelism that can effectively use the CPU as well as the GPUs. For a long time CUDA~\cite{cuda} has been the first choice for GPU programming. Developed by NVIDIA, CUDA provides an API to program GPUs that can be used in applications written in C/C++ or Fortran. HIP~\cite{HIP}is AMD's proprietary GPU programming environment. Although CUDA and HIP offer great performance for parallel applications, they often require programmers to rewrite their programs entirely and are platform specific.

As more vendors are entering the GPU market, portable parallel programming methods are required so that application programmers can run codes on diverse heterogeneous systems, such as Summit and Frontier, without massive re-engineering. Directive-based parallel programming models OpenMP~\cite{openmp08} and OpenACC offer an approach that allows users to annotate their serial code in a more straightforward manner and produce parallel versions of their applications that will run on many different architectures.
%OpenMP \cite{openmp08} is one of the options available to application developers to realize performance portability across different architectures. 

In preparation for the release of ORNL's Frontier and other US Department of Energy (DOE) funded systems, the DOE Exascale Computing Project (ECP) sought to prepare an Exascale software stack to ensure that mission-critical applications are able to embrace the potential performance boosts offered by newer generations of hardware. OpenMP, a parallel programming library, is one component of this software stack. More features that are valuable to developers continue to be added to the OpenMP specification. The objective of ECP's Scaling OpenMP via LLVM for Exascale Performance and Portability (SOLLVE) team is to ensure OpenMP is compatible with the unique software and hardware requirements of exascale computing and to enable seamless migration of applications to the novel computing system.

As of May 2022, the compilers that offer support for OpenMP offloading features (specification versions 4.0 and later) are AMD, Flang, GNU, HPE, IBM, Intel, LLVM, NVIDIA, and Siemens. While this list of compilers that support offloading with OpenMP is significant, there are far less compilers that have continued to expand their OpenMP implementations for versions 4.5, 5.0, and 5.1. According to the OpenMP website~\cite{compilers}, the only compilers that have any coverage of OpenMP 5.0 offloading features are AMD, GNU, HPE, Intel, LLVM, Siemens, and NVIDIA.

OpenMP 5.0 was released in November 2018 and intro- duced a wide variety of improvements on heterogeneous target offload and host based features. One new addition, the requires directive, allows the programmer to request features from the implementation that must be supported to enable proper execution of kernels in a given computation unit. Of these features available for enforcement, reverse offload and unified shared memory prove to the most valuable as they enable on-host execution initiated from the offload device and utilization of a shared memory space between devices, respec- tively. Another important feature released in OpenMP 5.0 is the declare mapper directive. The declare mapper directive now allows the creation of user-defined mappers to avoid ambiguities that can arise between explicit and implicit mapping of variables as well as the ability to map members of a struct or class.

Results and discussions entail evaluation of compilers' current status of stability and maturity of implementations, types of errors, and discussions that have led to the language committee revisiting the verbiage used in the specification.

\label{sec:introduction}

\section{Background and Motivation}
\subsection{OpenMP}
OpenMP Specification provides an Application Program Interface (API) to allow programmers to develop threaded parallel codes on shared memory systems. 
The OpenMP directives or \texttt{pragmas} are understood by OpenMP aware compilers while other compilers lacking OpenMP support are free to ignore them. Usually a flag such as \texttt{-fopenmp} is required at compile time to activate OpenMP recognition and processing by the compiler. Along with compiler directives, OpenMP also provides library routines and environment variables for explicit control. 
The OpenMP \texttt{parallel} directive generates parallel threaded code where the original thread becomes thread “0”. 
The new league of threads share resources of the original thread and the specific data-sharing attributes of variables can be specified based on usage patterns of the application. A basic usage example of the \texttt{parallel} directive is provided in the code-snippet below.

\begin{lstlisting}[language=C, caption=Simple C program using OpenMP for matrix-matrix addition, label=lst:ompcpu]
int A[N][N], B[N][N], C[N][N];
// initialize arrays
#pragma omp parallel for
  for (int i = 0; i < N; ++i) {
    for (int j = 0; j < N; ++j) {
      C[i][j] = A[i][j] + B[i][j];
    }
  }
\end{lstlisting}

\subsection{Offloading to Devices}

OpenMP device directives such as \texttt{target} provide mechanisms for an OpenMP program to offload parallel code and data to \textit{target devices}.
%OpenMP offers three levels of parallelism (\textit{teams}, \textit{threads}, and \textit{simd lanes}), but typical devices provide only two levels of parallelism; Intel CPUs offer thread and simd level parallelism, NVIDIA GPUs provide thread-block and thread level parallelism, and AMD GPUs provide work-group and work-item level parallelism. 
%Therefore, different OpenMP compilers can choose different parallelism mapping depending on target devices.

OpenMP offers three levels of parallelism (\textit{teams}, \textit{threads}, and \textit{SIMD lanes}), but existing devices may provide different levels of parallelism (typically two or three levels), and  different OpenMP implementations can choose different parallelism mapping for the same target device. Table~\ref{table:CM} lists the equivalence in terminology across different vendors and OpenMP.
For example, many of the existing OpenMP compilers largely ignore SIMD clauses when targeting GPUs (mapping OpenMP teams to GPU thread blocks and OpenMP threads to GPU threads), but typical GPUs also expose thread-scheduling units, such as \textit{warps} in NVIDIA GPUs and \textit{wavefronts} in AMD GPUs, and thus other OpenMP compilers may choose to provide fine-grained three level parallelisms (e.g., OpenMP teams to GPU thread blocks, OpenMP threads to GPU warps, and OpenMP SIMD lanes to GPU threads).
When targeting CPUs, however, the SIMD clause may play an important role, and most existing OpenMP compilers exploit the SIMD-level parallelism by mapping OpenMP threads to CPU threads and OpenMP SIMD lanes to CPU SIMD lanes.
But the applicability of the SIMD parallelism largely depends on the compiler's vectorization capability, and it is still implementation-defined how to map the three level OpenMP parallelisms to CPU parallelisms.
Therefore, this work will primarily focus on the functional portability of the existing OpenMP implementations.

OpenMP provides a relaxed-consistency, shared-memory model for a given device, which allows all OpenMP threads to access the device memory to store and retrieve variables.
In the OpenMP device data environments, each device has its own device data environment, which may or may not share storage with other devices.
OpenMP device directives offer various data-mapping options (via \texttt{map}) to specify how an original variable is mapped from the current task’s data environment to a corresponding variable in the target device data environment.

\subsection{New 5.X Features}

As newer architectures continue to evolve, so does the feature requirements of parallel applications. To accommodate these needs, the OpenMP ARB continues to add new features to the specification. 
One of the more intriguing features that was added to the 5.0 specification is \texttt{metadirective}, which allows a program to run different variants of an OpenMP directive as determined by a conditional statement. The metadirective provides the \texttt{when} clause, which receives arguments like \texttt{arch} (architecture) and \texttt{isa} (instruction set architecture). 
A common use case for this directive would be when the architecture is NVIDIA or \texttt{when(arch==nvidia)} we can call an OpenMP directive, say \texttt{\#pragma omp target}. 
When this condition is not met, we can instead define a default behavior such as \texttt{\#pragma omp parallel}. 
In 5.1, the \texttt{error} directive and \texttt{nothing} directive were added specifically for usage with the metadirective clause, and enable run time errors or non-action behaviors to occur when a condition in the \texttt{when} clause is not met.

%The \texttt{declare target} directive, also introduced in 5.0, allows the user to explicitly ensure that procedures and global variables can be accessed on a device. A key functionality of this new directive is allowing a user to create only device version, only host version, or both versions of a function that they wish to be included in the device memory. Functionality induced by this clause is somewhat similar to \texttt{metadirective} in that a user can make a host only or device only version of a function or global variable accessible. However, \texttt{metadirective} offers the added bonus of triggering different behavior based on a conditional statement.
OpenMP 5.0 was released in November 2018 and it introduced a wide variety of improvements for heterogeneous target offload and host based features. A new addition, the \texttt{requires} directive, allows the programmer to request features from the implementation that must be supported to enable proper execution of kernels in a given computation unit. 
Of these features available for enforcement, reverse offload and unified shared memory prove to the most valuable as they enable on-host execution initiated from the offload device and utilization of a shared memory space between devices, respectively. 
Another important feature released in OpenMP 5.0 is the \texttt{declare mapper} directive. 
The \texttt{declare mapper} directive now allows the creation of user-defined mappers to avoid ambiguities that can arise between explicit and implicit mapping of variables as well as the ability to map members of a struct or class.

The \texttt{declare target} directive, which was initially introduced in 4.0, allows the user to explicitly ensure that procedures and global variables can be accessed on a device.
The 5.0 specification extends this directive with additional functionalities and clauses.
For example, the new clause, \texttt{device\_type} allows a user to create only device version, only host version, or both versions of a function that they wish to be included in the device memory. Functionality induced by this clause is somewhat similar to \texttt{metadirective} in that a user can make a host only or device only version of a function or global variable accessible. 
However, \texttt{metadirective} offers the added bonus of triggering different behavior based on a conditional statement.

Many of the features introduced in 5.0, such as \texttt{metadirective} and \texttt{requires}, are implementation-dependent, meaning compiler vendors have some variability in the manner by which they choose to implement these features. 
The \texttt{requires} directive inherently requests that an implementation must be able to provide a certain behavior in order to compile and run a program correctly. 
The certain behaviors that can be requested or `required' by the programmer are reverse offloading, unified address, unified shared memory, atomic default memory ordering, and dynamic allocators. 
A user can request reverse offloading using \texttt{\#pragma omp requires reverse\_offload} at the top of their program. If the implementation does not have support for this feature, the program will either ignore the \texttt{requires} statement and issue a compiler warning or rather issue a compile error.

The \texttt{declare variant} directive, again, can be utilized to achieve a similar functionality as the \texttt{metadirective} and 
\texttt{declare target} directive. Utilizing the same context-selector-specification field as the \texttt{metadirective}, \texttt{declare variant} can call a different version of a provided base function, based on the context or conditional statement with which the directive is associated.

OpenMP 5.1, released in November 2020, introduces features such as the \texttt{assume}, \texttt{nothing}, \texttt{scope}, 
\texttt{interop} directives, loop transformation constructs, new modifier clauses that extend the \texttt{taskloop} construct, newer support for indirect calls to the device version of a function in target regions, amongst others.  

OpenMP 5.2 was released in November 2021 and continues to add onto the previous OpenMP developments. OpenMP 5.2 specifically made improvements in its memory allocators, use of Fortran PURE procedures, and use of the \texttt{scope} construct. OpenMP 5.2 also includes simplified unstructured data offload use, extended support of user-defined mappers, more consistent \texttt{linear} clause, and refined OpenMP directives syntax.

\label{sec:background}

%\section{Motivation}

%The need to validate and verify the compiler implementations of OpenMP features becomes more important than ever as vendors implement the newer and more complex features discussed above. From the ECP perspective, there is an urgent need for vendor agnostic verification testsuite as the Frontier super computing system is being made available for use by application and software developers. The main objective of the testsuite is not comprehensive coverage of the new OpenMP features, but a comprehensive coverage of the new features that are important to HPC application developers (mainly in the DOE and ECP space). 

%\label{sec:motivation}

\section{SOLLVE Validation and Verification Suite}
The SOLLVE Validation and Verification testsuite was built to provide open-source vendor agnostic feature tests for the latest OpenMP Specifications with focus on features of interest to applications. The process for collecting application input across ORNL and other DOE labs is outside the scope of this paper. Like most software projects, SOLLVE V\&V is aware that with vendor implementations and application changes, the importance of features may change over time and we perform regular checks and add missing tests/corner cases to the testsuite.
 
 \subsection{Test Creation Strategy}

Within every new release of the OpenMP specification, there is a section that details the differences between the most current version and its predecessor, which outlines all of the new features that are provided to users. Compiler developers from LLVM, GNU, and more take this differences list, or a similar list potentially provided by the OpenMP ARB, and formulate a to-be-implemented list that is typically hosted on their website. For LLVM, each of the features will have a status associated with it that describes the progress made thus far in implementing: either unclaimed, worked on, mostly done, not upstream, or done. 
Our development of feature tests is dictated first by the ECP application needs. Through our interactions with the Application Development (AD) teams, a priority list of the most desired new features was created. 
%Because the role of our test suite is to ensure that OpenMP features are implemented correctly, we first choose features that are 'done', and proceed to create tests that evaluate several aspects of the new feature. 
When writing a new test for an already implemented feature, the usability of the feature that is outlined by the specification is analyzed. Then, the potential combinations of options that may be presented are outlined. For example the \texttt{default} clause which has options for \texttt{shared}, \texttt{none}, as well as \texttt{firstprivate} \& \texttt{private} which were introduced in OpenMP 5.1. 
For this example, there would need to be two tests to encapsulate the full functionality of this new feature. Lastly, careful attention is paid to the `restrictions' section of each feature, to ensure that the test being written does not violate boundaries that have been outlined by the OpenMP Specification. In the case of OpenMP features that do not have implementations, generating a brand new test that is both syntactically correct and accurate can prove difficult. 
%Again this test creation process starts with consulting the OpenMP specification for the appropriate release. The test then begins its development cycle. 

In either case it is ensured that the test meets all conditions and restrictions noted in the specification and provides adequate error checking in case of failure. During this stage of development it is common to receive feedback from other collaborators on how to approach or improve the test, especially on new tests that do not have implementations. After the test has been written it enters a review process. This process includes having each test independently verified by two other collaborators, internal or external to our team. 
%A visual flowchart of the workflow is depicted in Figure \ref{fig:flowchart1}. 
Other interested parties, including members form the OpenMP community or ARB, are welcome to provide their input in the form of Github issues or pull-requests.
%\begin{figure}
%\centering
%\scalebox{.41}{\includegraphics{test_creation_flowchart.png}}
%\caption{A typical test creation workflow}
%\label{fig:flowchart1}
%\end{figure}

\subsection{OpenMP 5.x+ Feature Coverage}

As of this writing the V\&V testsuite includes 258 5.0 tests, 53 5.1 tests \& 6 5.2 tests. Since the primary focus of the SOLLVE project is development of OpenMP support in Clang, the testsuite has more coverage in C/C++. The 5.0 coverage includes a mix of Fortran \& C versions while the majority of our 5.1 tests are coded in C. Tests have been written for a vast majority of 5.1 features. Test priority related to 5.1 features are based primarily based on the needs of SOLLVE application developers, starting with high-priority, implemented features, and working our way to low-priority non-implemented features. For the new features introduced in OpenMP Specification a) 5.0 SOLLVE V\&V testsuite has 100\% coverage for C/C++, 70\% coverage for Fortran, b) 5.1 Specification 85\% coverage for C/C++, 5\% coverage for Fortran, and c) 5.2 Specification we have 20\% overall coverage. As mentioned before, the objective is not to have feature tests for all the new features but to have tests that cover, in sufficient detail, the important features as indicated by the ECP\/DOE applications.

\subsection{Challenges}
\subsubsection{Testing Unimplemented Features}

Often times tests cases must be written for new OpenMP features that are not yet implemented by any of the major compiler vendors. This makes the OpenMP specification one of the only resources available to understand how the feature would work once implemented. This can lead to some issues when attempting to develop strong tests for new features. A prime example of this is the test case for the \texttt{nothing} clause extension of the \texttt{metadirective} construct.

Here is an example of a simple implementation:
\begin{lstlisting}[language=C, caption=Simple usage of the nothing clause with the metadirective, label=lst:ompcpu]
#pragma omp metadirective \
     when( device={arch("nvptx")}: nothing) \
     default( parallel for )
     {
        for (int i = 0; i < N; i++) {
           A[i] += 2;
        }
     }
\end{lstlisting}

At a high level, the metadirective provides a way to dynamically change what OpenMP constructs are rendered. In the example above, if the code is offloaded to an NVIDIA device then the \texttt{nothing} clause would be rendered. If not, it would default to a \texttt{parallel for} loop. At first glance this seems pretty intuitive. Our initial interpretation of the specification was that the \texttt{nothing} clause would simply negate the code. In other words, it would not run the for loop if the code was running on an NVIDIA device. Based off the specification itself and various examples this seemed to be correct. This then lead to the bigger question of how to properly test nothing.

This test was initially written by looking for any spawned threads, or signs that the array had been changed and the code had run despite the \texttt{nothing} clause. This seemed promising, but it was discovered the team's interpretation was not the same as the compiler implementation, as the specification was vague. The \texttt{nothing} clause when used outside of the metadirective implies that OpenMP would ignore the \texttt{pragma} statement. However, the code would still run in serial. This meant the initial version of the test was wrong and needed to be amended.

Ultimately the test was reworked to determine if the metadirective had properly used the nothing directive by checking if the code was running in parallel instead of just checking for threads by leveraging the runtime \texttt{omp\_in\_parallel} function. This function would only return 1 if the code is running in parallel. In the example above, that would mean it would only be 1 if the nothing directive was not rendered properly. This ended up being a much more robust way to test the \texttt{nothing} clause with metadirectives and was utilized in the final version.

The \texttt{nothing} metadirective test demonstrates the challenge of the SOLLVE team's interpretation of a test compared to a compiler vendor's interpretation, and illustrates the need for heavy review and rewriting of code.

\subsubsection{Unclear Specification}

\begin{figure}
\includegraphics[width=0.5\textwidth,]{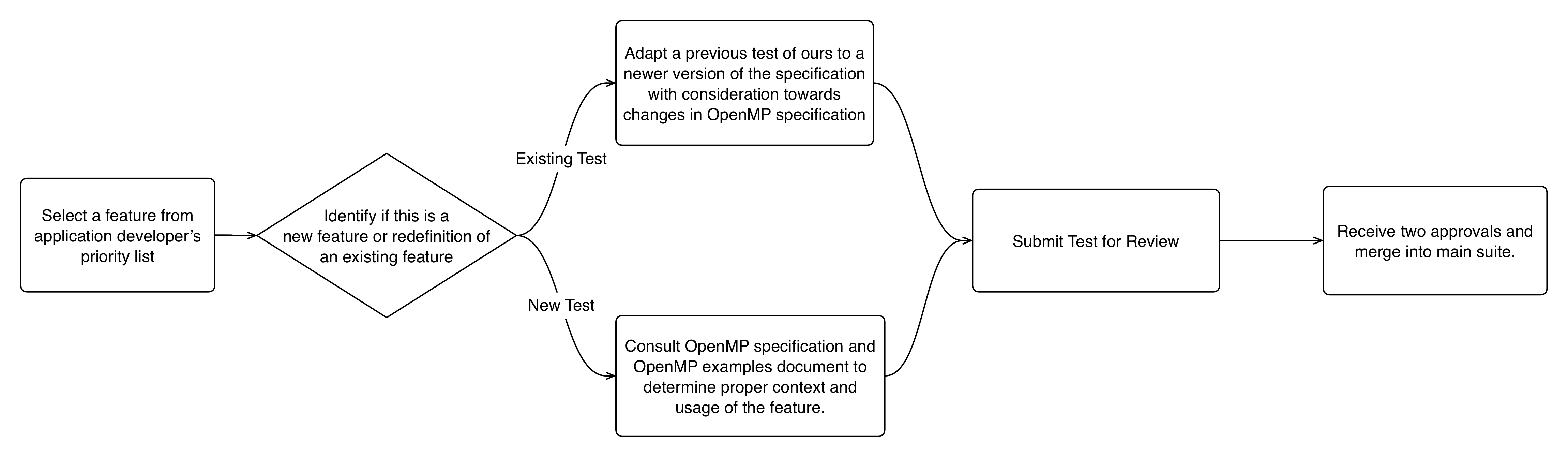}\caption{Process flow demonstrating the typical test creation
process.}
\label{fig:processflow}
\end{figure}

Another example of confusion relating to interpretation of the specification arose when writing a test case for the \texttt{has\_device\_addr} clause, added to the \texttt{target} construct in OpenMP 5.1. The description of this clause states: "The \texttt{has\_device\_addr} clause indicates that its list items already have device addresses and therefore they may be directly accessed from a target device."\cite{omp_target} While this may seem straight-forward, the purpose of this is relatively unclear. Questions arise, such as whether these list items be mapped first, and then marked as on the device? If the list items are already on the device, what is the benefit of listing them under the clause? How is it ensured that the list items are not unmapped at the end of a device region so that they remain when utilizing the clause? The difference between \texttt{use\_device\_addr} and \texttt{has\_device\_addr} is not clearly stated in the specification.

Furthermore, this clause was not listed on the OpenMP examples document.\cite{ompexamples} This document is often used by the SOLLVE team to assist in creation of tests that have no yet been implemented, as that document is the only official resource supported by the OpenMP ARB which shows the intended purpose and proper syntax of a new feature.

\begin{lstlisting}[language=C, caption=Example of has\_device\_addr directive, label=lst:ompcpu]
#pragma omp target enter data map(to: x, arr)
  #pragma omp target data use_device_addr(x, arr)
  #pragma omp target map(from:second_scalar_device_addr, second_arr_device_addr) has_device_addr(x, arr) 
  {
    second_scalar_device_addr = &x;
    second_arr_device_addr = &arr[0];
  }
#pragma omp target exit data map(release: x, arr)
\end{lstlisting}

The agreed-upon solution for this test arose only after having community-driven detailed discussion on the directive's purpose and the implicit mapping of target directive. It was decided that a \texttt{target enter data map} should be used to ensure variables are properly mapped to the device. Then, the \texttt{use\_device\_addr} and \texttt{has\_device\_addr} can be used in tandem to ensure the variables maintain their device addresses in the \texttt{target} region. 
%% Try not to use "we" ^^

\label{sec:vnv}

\section{Results and Discussion}
\subsection{Results from Summit}% - A Pre-Frontier System}
%SP: It is a supercomputer in it's own right. We do not have to associte it with Frontier
The following subsections shows results of GNU, LLVM and NVHPC compilers and their maturity over time, on Summit. 

\subsubsection{GNU Maturity Over Time}
\begin{figure}
\includegraphics[width=0.5\textwidth]{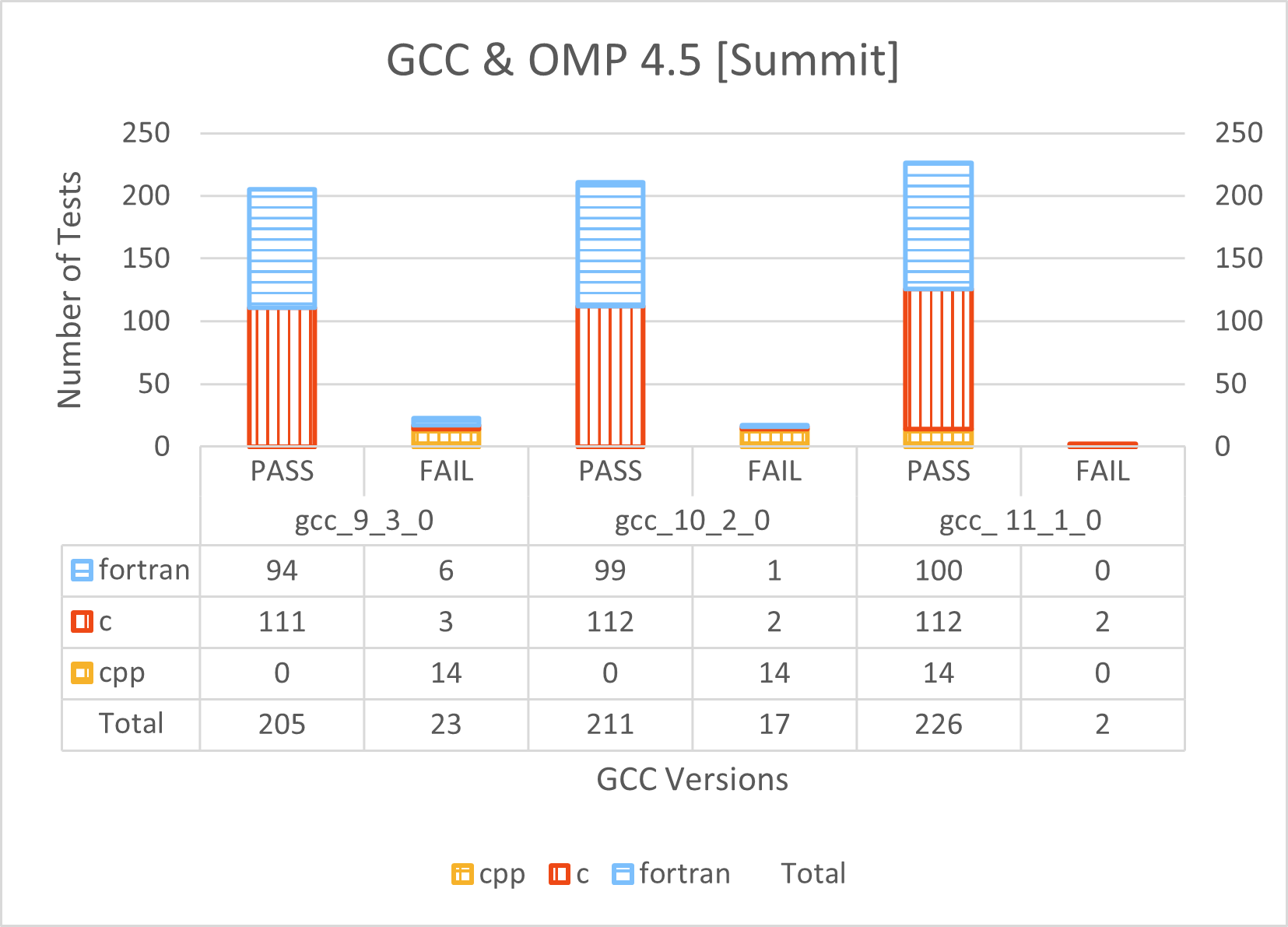}\caption{OpenMP 4.5 tests with GCC on Summit.}
\label{fig:gcc_4_5}
\end{figure}

\begin{figure}
\includegraphics[width=0.5\textwidth]{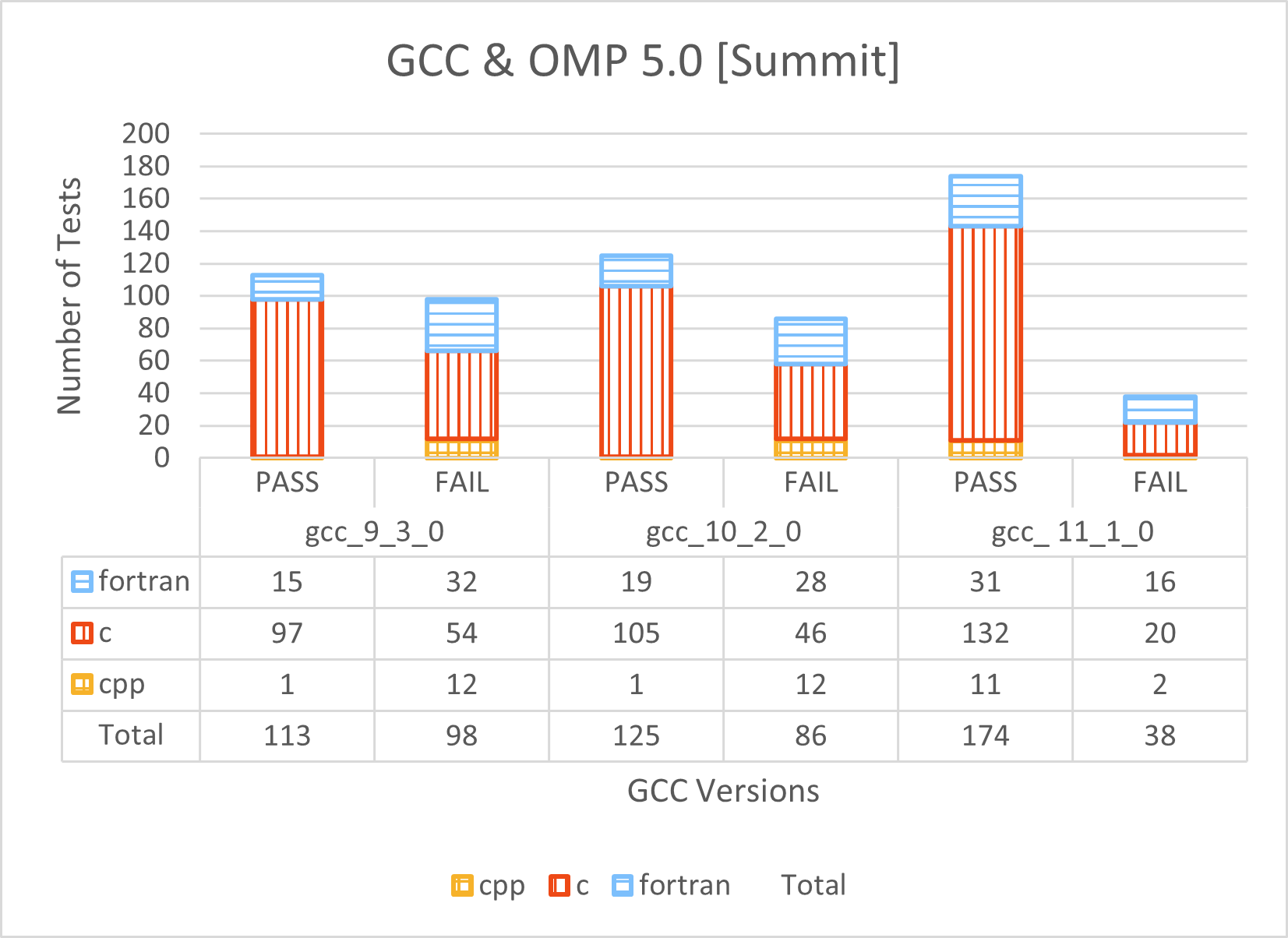}\caption{OpenMP 5.0 tests with GCC  on Summit.}
\label{fig:gcc_5_0}
\end{figure}

\begin{figure}
\includegraphics[width=0.5\textwidth]{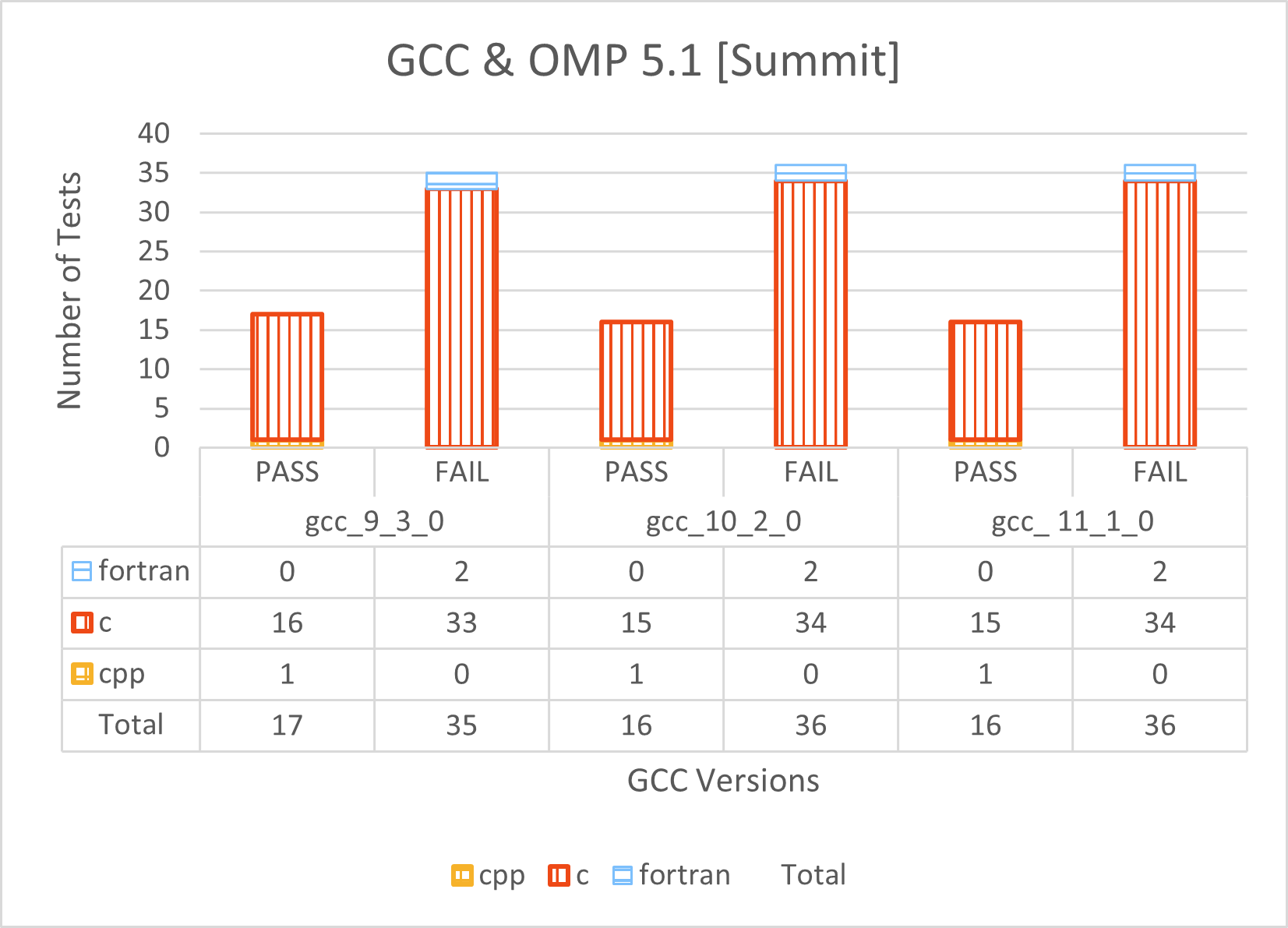}\caption{OpenMP 5.1 tests with GCC on Summit.}
\label{fig:gcc_5_1}
\end{figure}

For the GNU results shown in Figures \ref{fig:gcc_4_5}, \ref{fig:gcc_5_0} \& \ref{fig:gcc_5_1}, we only utilize stable releases of the compiler that are made available on OLCF's Summit supercomputer.
Regarding the GNU compilers, gcc and g++, there are seemingly a linear increase in support for both 4.5 and 5.0 features in OpenMP across major version releases. It is also important to note that GNU-11.2.0 is the first version of the compiler that supports features described in the OpenMP 5.1 specification. Version 12 Release of GNU supports far more 5.1 features than version 11.2.0, so any users attempting to utilize OpenMP 5.1 and 5.2 features with the GNU compiler should aim to use GNU version 12. Results in Table \ref{table:GCC_reg} show a list of tests that have passed and failed over a set of GCC compiler versions they have been tested on. For OMP 5.0 tests for \texttt{loop reduction and/or device} passes on GCC version 9.3.0 but fails in the next two versions i.e. 10.2.0 and 11.1.0.

\begin{figure}
\includegraphics[width=0.5\textwidth]{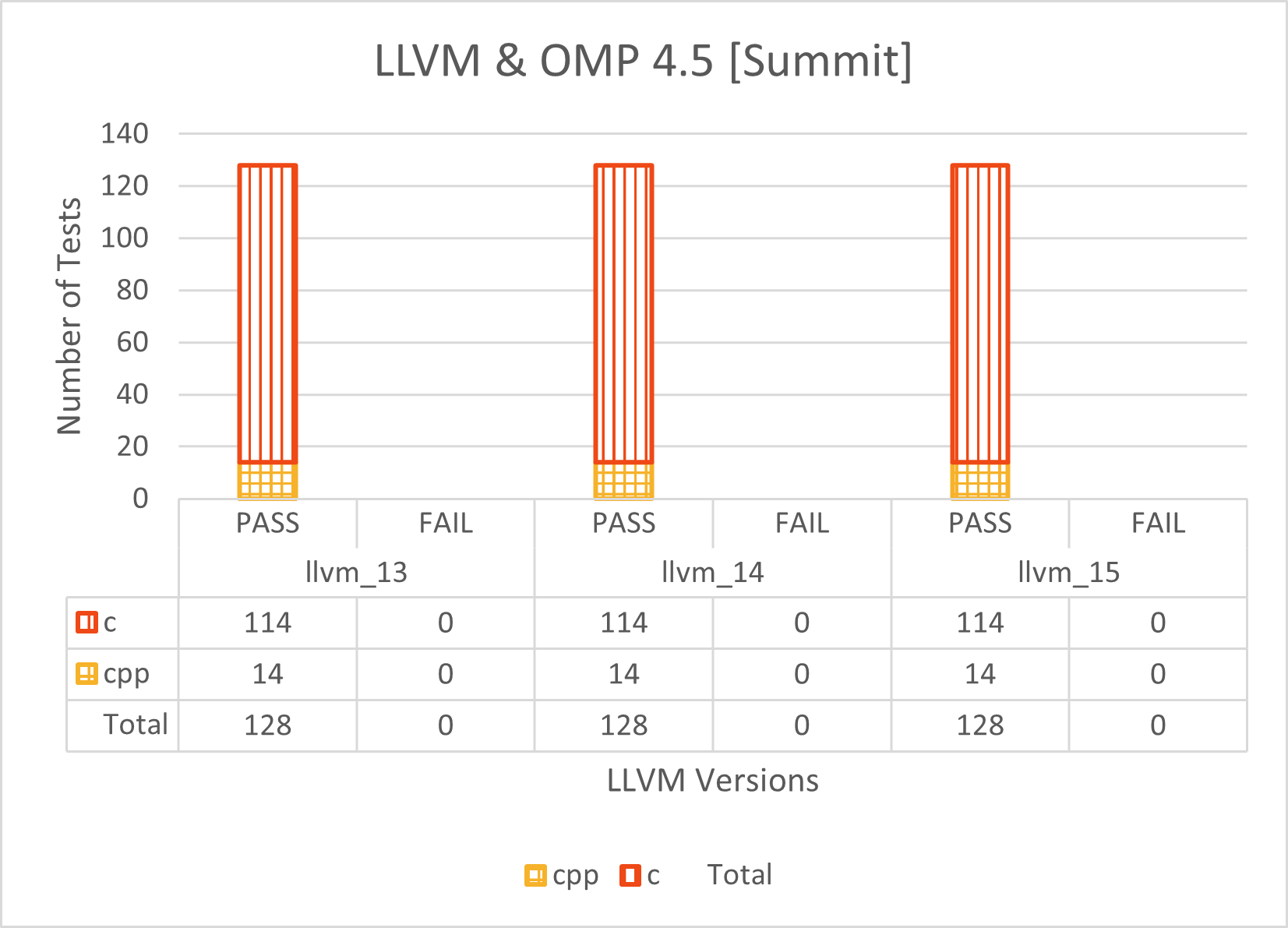}\caption{OpenMP 4.5 tests with LLVM on Summit.}
\label{fig:llvm_4_5}
\end{figure}

\begin{figure}
\includegraphics[width=0.5\textwidth]{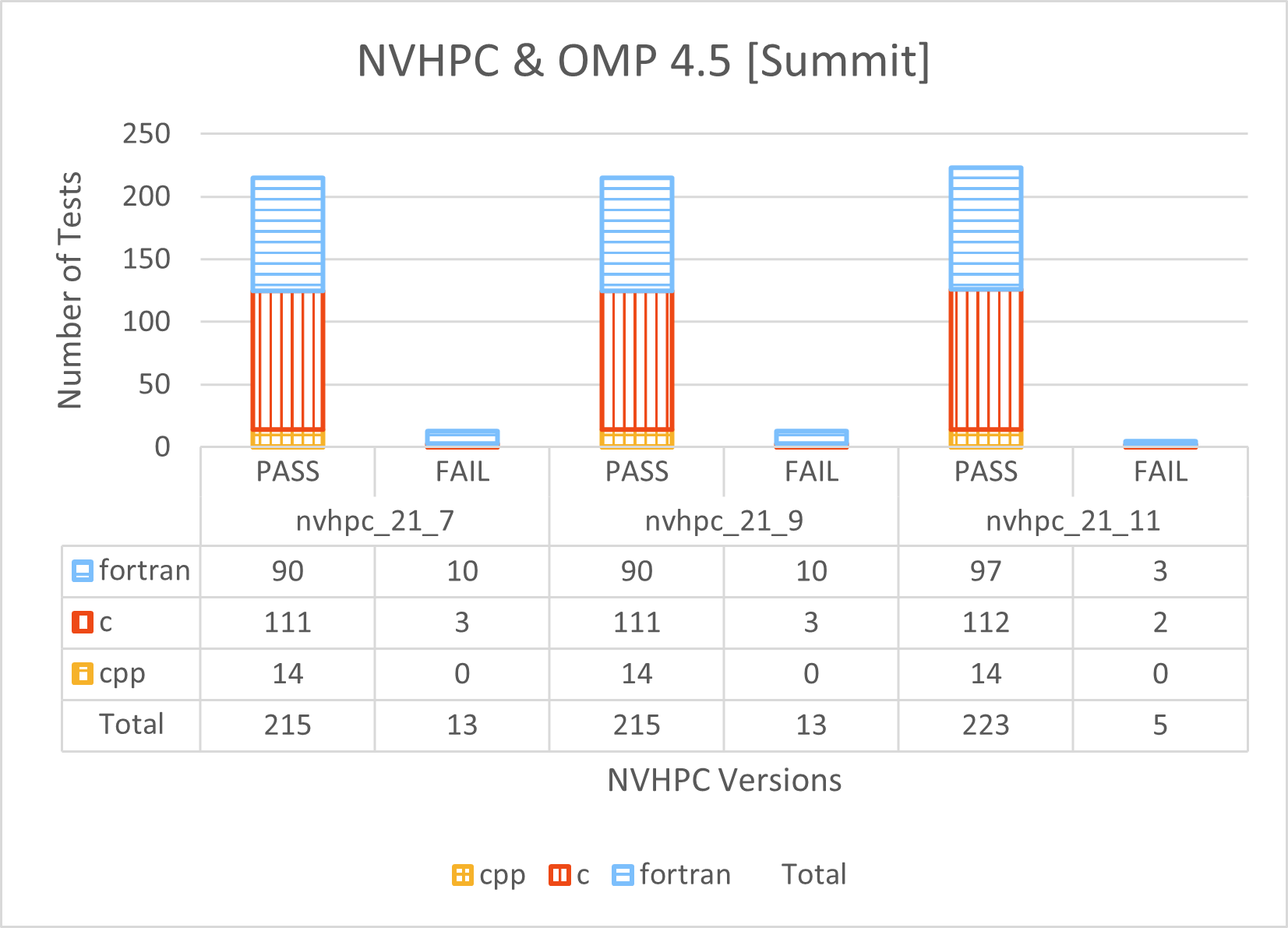}\caption{OpenMP 4.5 tests with NVHPC on Summit.}
\label{fig:nvhpc_4_5}
\end{figure}

\begin{figure}
\includegraphics[width=0.5\textwidth]{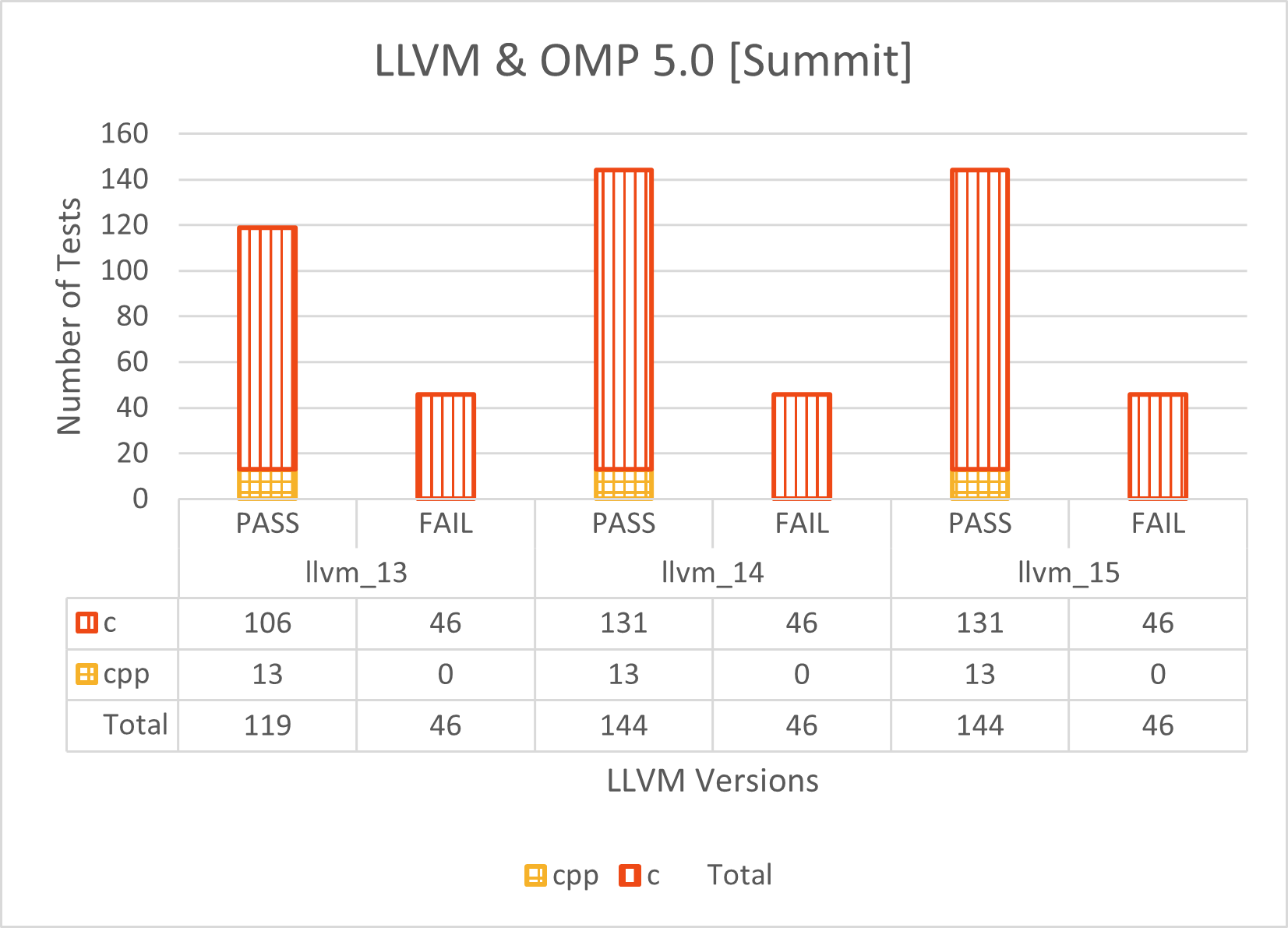}\caption{OpenMP 5.0 tests with LLVM on Summit.}
\label{fig:llvm_5_0}
\end{figure}

\begin{figure}
\includegraphics[width=0.5\textwidth]{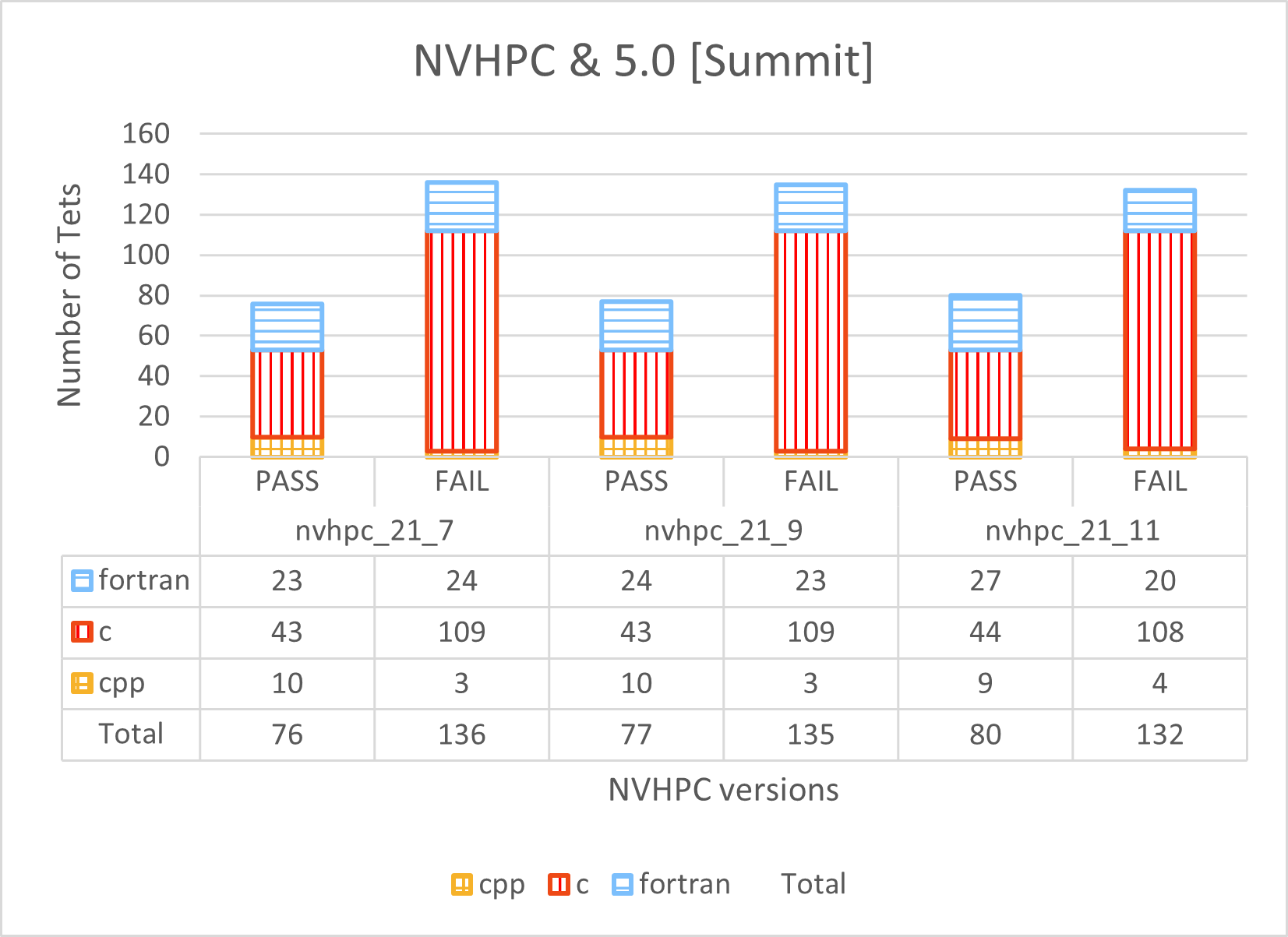}\caption{ OpenMP 5.0 tests with NVHPC on Summit.}
\label{fig:nvhpc_5_0}
\end{figure}

\begin{figure}
\includegraphics[width=0.5\textwidth]{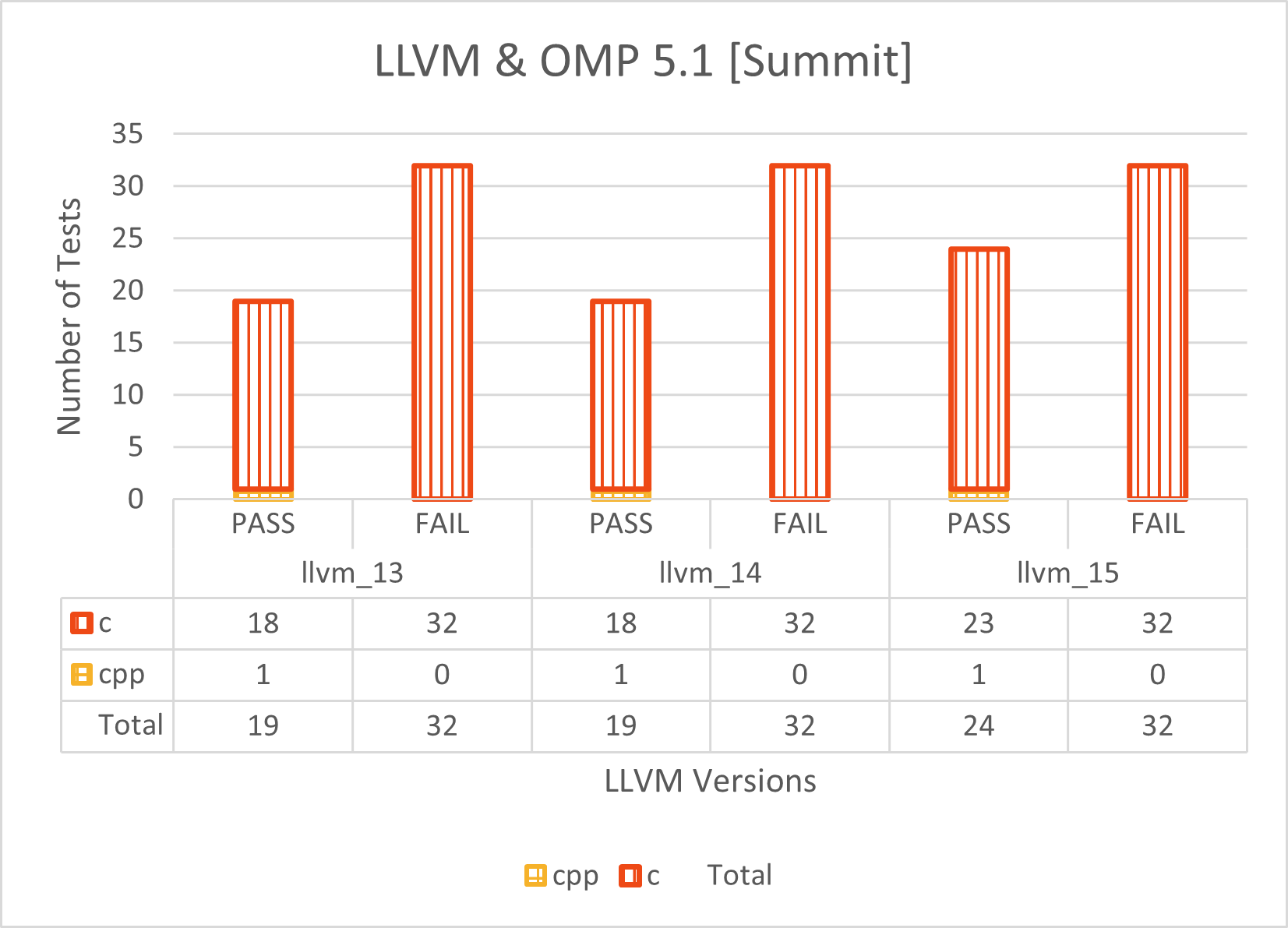}\caption{OpenMP 5.1 tests with LLVM on Summit.}
\label{fig:llvm_5_1}
\end{figure}

\begin{figure}
\includegraphics[width=0.5\textwidth]{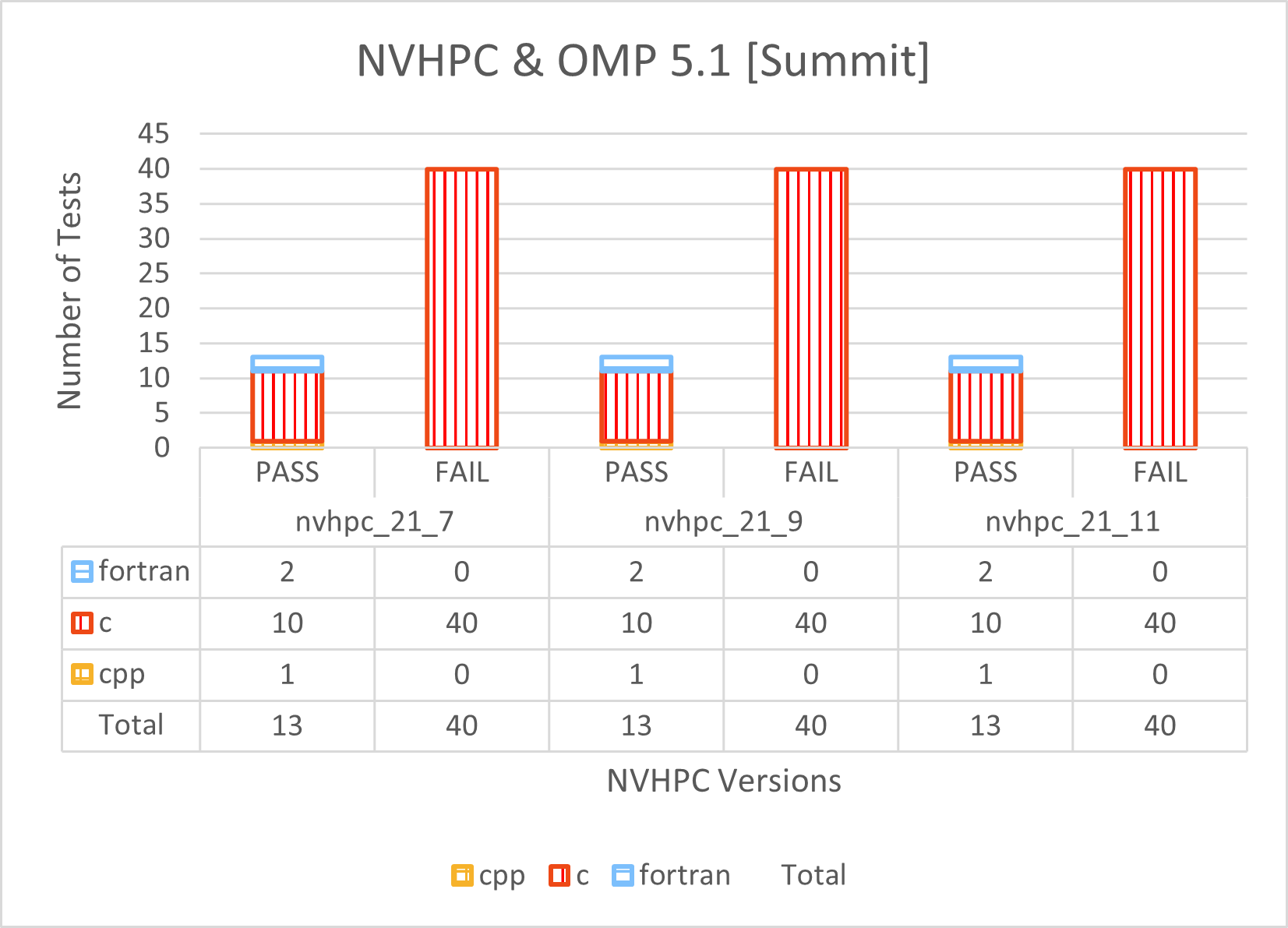}\caption{OpenMP 5.1 tests with NVHPC on Summit.}
\label{fig:nvhpc_5_1}
\end{figure}

\begin{figure}
\includegraphics[width=0.5\textwidth]{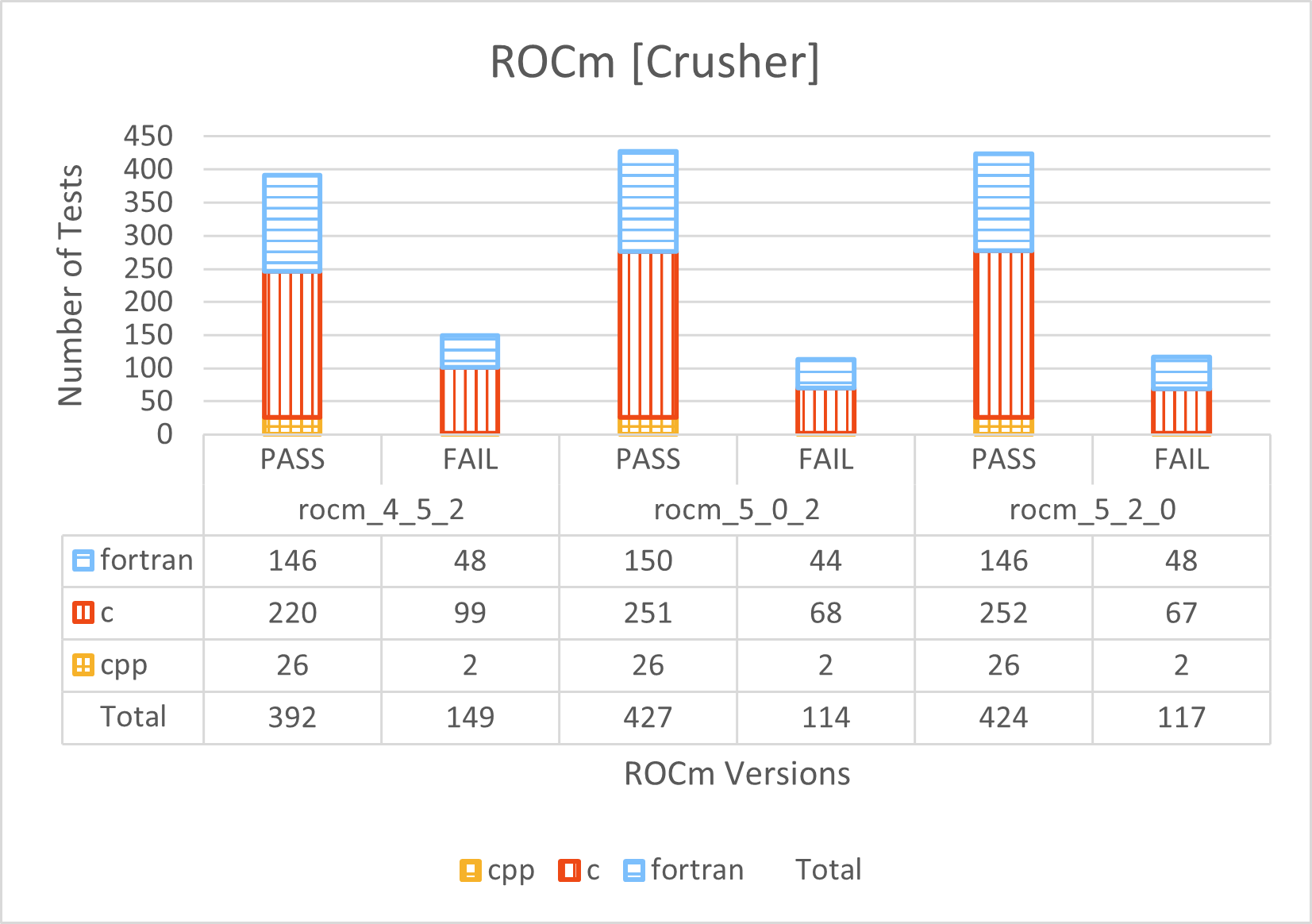}\caption{ OpenMP tests with ROCm on Crusher.}
\label{fig:rocm}
\end{figure}
\begin{figure}
\includegraphics[width=0.5\textwidth]{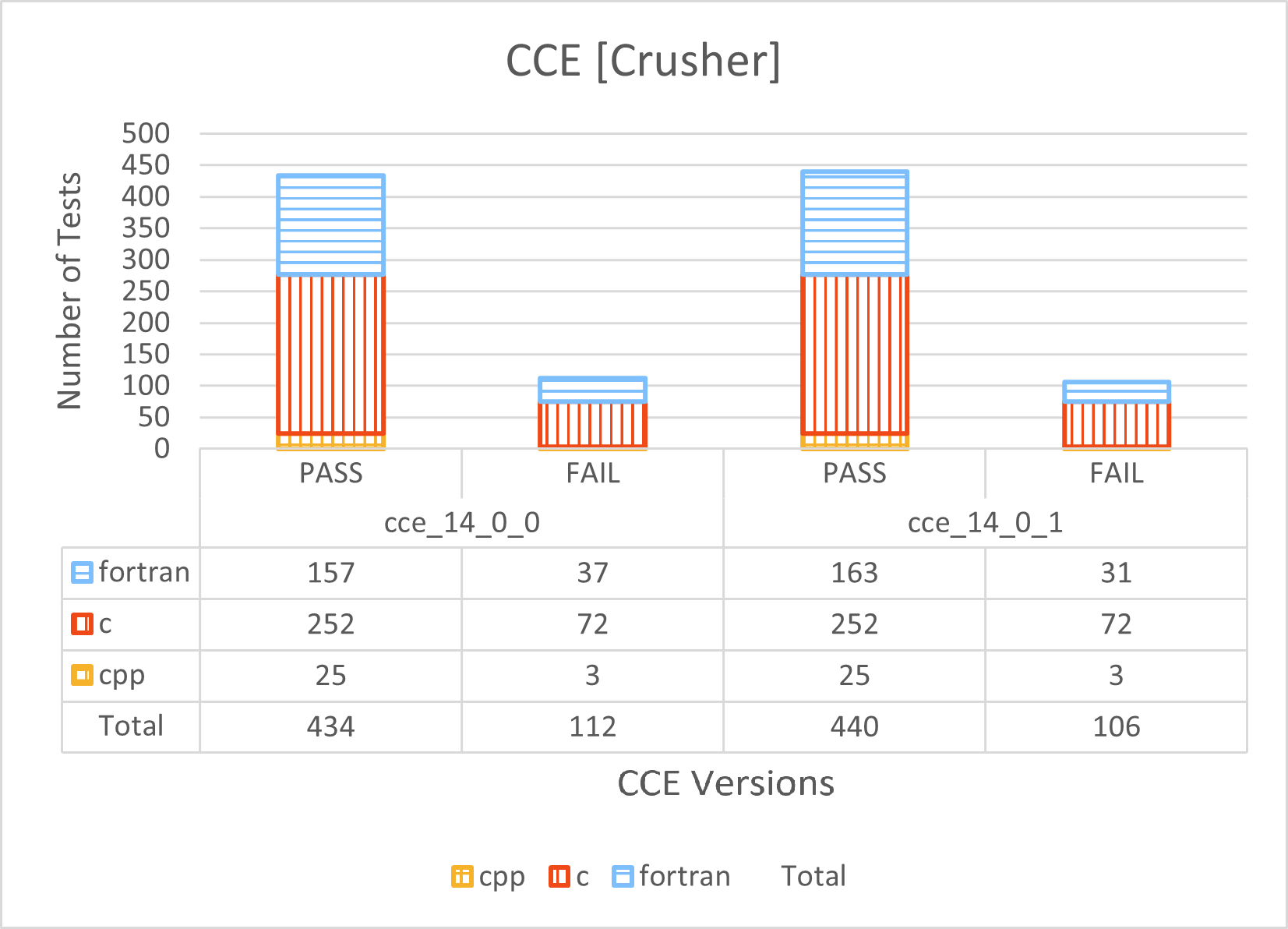}\caption{OpenMP tests with CCE on Crusher.}
\label{fig:cce}
\end{figure}

\subsubsection{LLVM Maturity Over Time}
The results presented in Figures \ref{fig:llvm_4_5}, \ref{fig:llvm_5_0} \& \ref{fig:llvm_5_1}, are for Clang and Clang++ using the stable releases of LLVM-13, LLVM-14 and the LLVM-15 developmental release available on the OLCF's Summit supercomputer. 
It is important to note that LLVM provides an \texttt{-fopenmp-version} flag that allows you to inform the compiler which specification version of OpenMP you would like to compile for. 
This is vital for testing various implementations of features, as many times features will get redefined by new versions of the specification. 
For example, the \texttt{master} construct in OpenMP 5.0 was renamed to \texttt{masked} in OpenMP 5.1. The important thing to note here is that LLVM continues to introduce more and more OpenMP 5.1 features. There are some rollbacks in 4.5 and 5.0 which could be due to features being 'completed' and then reopened for further investigation, becoming 'partial'. Results in Table \ref{table:LLVM_reg} show a list of tests that have passed and failed over a set of LLVM compiler versions they have been tested on. For OMP 5.0 test written for \texttt{master loop device} passes on LLVM version 13 but fails in the next two versions i.e. 14 and 15. Similar trends are seen for five other OMP 5.0 tests while the OMP 5.0 test for \texttt{reverse offload} only fails on LLVM version 15, which is the latest tested version.

\begin{table}[h!]
\centering
\begin{tabular}{*{5}{|c}|}
\hline
    Test Name & OMP Ver &gcc 9.3.0 & gcc 10.2.0  & gcc.11.1.0\\ \hline
    test\_loop\_reduction & 5.0 &Pass & Fail & Fail \\
    \_and\_device.c & & &  &  \\ \hline
    test\_loop\_reduction & 5.0 & Pass & Fail & Fail \\ 
    \_or\_device.c & &  &  &  \\\hline
\end{tabular}
\caption{Inconsistencies of tests passing and failing across different GCC versions}
\label{table:GCC_reg}
\end{table}

\begin{table}[h!]
\centering
\begin{tabular}{*{5}{|c}|}
\hline
    Test Name & OMP Ver & llvm\_13 & llvm\_14  & llvm\_15\\ 
    \hline
    test\_master\_taskloop & 5.0 & Pass & Fail & Fail \\
    \_device.c & &  &  &  \\
     \hline
    test\_master\_taskloop\_ & 5.0 & Pass & Fail & Fail \\ 
    simd\_device.c & &  &  &  \\
    \hline
    test\_parallel\_master & 5.0 &Pass & Fail & Fail \\
    \_device.c & &  &  &  \\
   \hline
    test\_parallel\_master\_ & 5.0 &Pass & Fail & Fail \\ 
    taskloop\_device.c & & &  &  \\\hline
    test\_parallel\_master & 5.0 & Pass & Fail & Fail \\
    \_taskloop\_simd\_device.c & & &  &  \\ \hline
    test\_requires\_reverse & 5.0 & Pass & Pass & Fail \\ 
    \_offload.c & & &  &  \\\hline
    test\_target\_task\_depend & 5.0 &Pass & Fail & Fail \\
    \_mutexinoutset.c & & &  &  \\ \hline
\end{tabular}
\caption{Inconsistencies of tests passing and failing across different LLVM versions}
\label{table:LLVM_reg}
\end{table}

\subsubsection{NVHPC Maturity Over Time}
The results collected in Figures \ref{fig:nvhpc_4_5}, \ref{fig:nvhpc_5_0} \& \ref{fig:nvhpc_5_1}, regarding NVHPC are on OLCF's Summit supercomputer system. We targeted the last three stable releases of the NVIDIA HPC compiler suite including the latest, 21.11. For these results, it is important to note that while coverage for 4.5 is about complete, acceleration of coverage for 5.0 has not increased quickly over the last few releases. Additionally, only 13 of the 53 features that we have written tests for OpenMP 5.1 for are supported. Results in Table \ref{table:NVHPC_reg} show a list of tests that have passed and failed over a set of NVHPC compiler versions they have been tested on. For OpenMP 4.5 version, tests written for \texttt{target teams distribute for if parallel modifier} passes using NVHPC version 21.7 but fails in the next two versions i.e. 21.9 and 22.11. Similar trends can be seen with tests in OpenMP version 5.0.
%Here we can maybe cite a paper where NVIDIA shows better speed performance, such that we are not disparaging the nvhpc compiler entirely.

\begin{table}[h!]
\centering
\begin{tabular}{*{5}{|c}|}
\hline
    Test Name & OMP Ver & 21.7 & 21.9  & 21.11\\ \hline
    test\_target\_teams\_ & 4.5 &Pass & Pass & Fail \\
    distribute\_for\_ & & &  &  \\ 
    parallel\_for\_if\_ & & &  &  \\ 
    parallel\_modifier.c & & &  &  \\ \hline
    lsms\_triangular\_packing.cpp & 5.0 & Pass & Pass & Fail \\ 
    \hline
     test\_declare\_variant.F90  & 5.0 & Pass & Pass & Fail \\ 
     \hline
    test\_target\_ & 5.0 &Pass & Pass & Fail \\
    teams\_distribute\_parallel & & &  &  \\ 
    \_for\_collapse.c & & &  &  \\
      \hline
\end{tabular}
\caption{Inconsistencies of tests passing and failing across different NVHPC versions}
\label{table:NVHPC_reg}
\end{table}

\subsection{Results from Crusher - A Pre-Frontier System}
Here we share the evaluation of compiler implementations on Crusher. We evaluate AMD ROCm and Cray CCE compilers. 

\subsubsection{ROCm Maturity Over Time}
The results listed in Figure \ref{fig:rocm} are from the Pre-Frontier Crusher system and show 4 versions of AMD's developmental HPC ROCm compiler. Results show a leap in OpenMP implementation from version 4.5.0 to version 5.0.0, but minimal changes there onward. It is interesting to note that one C test now passed from version 5.1.0, but one Fortran test now fails. This, again, could be due to features requiring more investigation or the definition of features being changed in the newer versions of OpenMP. 

\subsubsection{CCE Maturity Over Time}
The results listed in Figure \ref{fig:cce} also show the Cray Compiling Environment (CCE) results on Crusher. The results only include CCE 14.0.0 \& CCE 14.0.1 versions, as the only other versions available on Crusher, 13.0.0 \& 13.0.2 do not work properly with OpenMP. 
These versions require dependencies from both ROCm 4 \& ROCm 5, which cannot be loaded at the same time. The results show decent performance, with around 80\% of tests passing for version 14.0.0, increasing slightly with 5 more Fortran tests passing in 14.0.1. 
It is interesting to note that C implementation for CCE \& ROCm compiler is nearly identical, but Fortran implementation on CCE is slightly better.

\subsection{Subset of OpenMP Supported By All Compilers}
The run-time and compile-time results for NVIDIA, LLVM, GNU, CCE, and ROCm on the Summit and Crusher supercomputing systems reveal a subset of OpenMP features supported by all of the aforementioned compilers. Also revealed, is a subset of OpenMP features not supported by any of the aforementioned compilers. In this section's discussion, a pass is only deemed a pass if it is such for each of the five compilers. Thus, a fail is only deemed a fail if it is such for each of the five compilers. Regarding discussion of Fortran in this section, we will be analyzing results for NVIDIA, GNU, CCE, and ROCm exclusively, as LLVM does not have support for OpenMP offloading on the Summit supercomputer. This point will be belabored in order to avoid any misrepresentation. For our C/C++ OpenMP 4.5 tests, we report 85.93\% tests pass and 0\% of tests fail across all five compilers. Moreover, for the Fortran OpenMP 4.5 tests, 74.50\% of these tests pass and 0\% of tests fail across GNU, NVIDIA, CCE, and ROCm. These results exemplify OpenMP 4.5's maturity as almost all of the OpenMP 4.5 features that tested portable across all of the five compilers for C/C++ and four of the compilers for Fortran. Equally as impressive is the fact 0\% of tests fail across the aforementioned compilers for C/C++ and Fortan, meaning that every feature tested in OpenMP 4.5 is supported by at least one compiler. Moving onward to OpenMP 5.0, we report 18.34\% of the C/C++ tests pass and 4.14\% fail across all five compilers. For the Fortran OpenMP 5.0 tests, we report a 16\% pass rate and a 23\% fail rate across NVIDIA and GNU. While OpenMP 5.0 is certainly not as portable as OpenMP 4.5, the results expose the fact that almost every OpenMP 5.0 feature we test is supported by at least one compiler. For the C/C++ OpenMP 5.1 tests, only 1.96\% of tests pass and 43.13\% of tests fail for C/C++ tests pass across all three compilers. Finally, the Fortran OpenMP 5.1 tests have a 0\% pass rate and a 50\% fail rate across NVIDIA and GNU.

%\subsection{Kernel performance across compiler vendors}

%Pick an application kernel or find plausible kernel from ECP apps that utilizes one of the 'show-stopper' features in OpenMP 5.0. Report on performance or at least conformance across compilers for said feature. Test on various next-gen systems (Spock, Tulip, pre-frontier, etc)
\label{sec:discussion}

\section{Impacting OpenMP Community, Vendors, OpenMP Specification, and Applications}
This section draws inferences from the lessons learnt via this project and highlights the impact of SOLLVE V\&V on on the different aspects related to OpenMP; mainly community, vendors, specification, and applications. 

%Though not comprehensive, in this section we provide samples of how the SOLLVE V\&V has an impact on the different aspects related to OpenMP; mainly community, vendors, specification, and applications. 
\subsection{Impact on the OpenMP Community}
With rapid development of the OpenMP Specification and with OpenMP Examples ~\cite{ompexamples} as the only OpenMP ARB sanctioned resource, application programmers and other users often refer to the SOLLVE V\&V tests to see how to utilize a new OpenMP features.
One of reasons for this is that OpenMP Examples ~\cite{ompexamples}, though an excellent resource, is not exhaustive and is often released after a given specification version has been around for sometime.
The SOLLVE V\&V also consists of some tests adapted from OpenMP Examples document, but recently, the examples document has been extended to include an \texttt{declare target} example based on the SOLLVE V\&V feature test for \texttt{device\_type(nohost)} based on the community discussion during the test creation.
Listing ~\ref{lst:ompdt} shows the skeleton testcase. Due to the fundamental OpenMP requirement that fallback execution of device constructs must be supported, using \texttt{device\_type(nohost)} for procedure \textit{target\_fun} on the \texttt{declare target} imposes an additional requirement not clearly mentioned in the specification. 
Since \texttt{nohost} implies that the procedure \textit{target\_fun} is made available only on the device, it needs to be a device variant for the procedure \textit{fun()}. 
This is needed to ensure that a host symbol for \textit{target\_fun} is not required to be present in the host environment in the case of host fallback.
Without the variant function, the use of \texttt{nohost} will result in a link time error due to the code generated for host execution of the target region.

\begin{lstlisting}[language=C, caption=Snippet from declare target directive test with device\_type nohost, label=lst:ompdt]
...
#pragma omp declare variant(target_fun) match(device={kind(nohost)})
void fun() {
  /*some work*/
} }
void target_fun(){
  /*some work*/
}
#pragma omp declare target enter(target_fun) device_type(nohost)
int main() {
  ...
  #pragma omp target
  {
    foo(); // calls the target_fun() on device or fun() in case of host fallback.
  }
  ...
\end{lstlisting}
%

%When creating a test for a feature that has no easy to find examples online, we run into situations where we may misinterpret the specification. Other times, we run into issues where the specification itself is not clear and could be interpreted many different ways. Either way, putting out a new test often generates discussions amongst people who monitor our test suite, and sometimes this discussion may result in submitting issues to the OpenMP Language Committee.

\subsection{Impact on Vendor Implementation}
%\subsection{Usage Error leads to Upstream Patch in GCC}
The test cases we create can also help expose missing aspects of a specific compiler's implementation. In one instance \url{https://github.com/SOLLVE/sollve_vv/issues/409}, we were developing a test to evaluate the new \texttt{metadirective} feature. The objective of the test was to check the use of context-selectors to determine which vendor provided the implementation, either AMD or NVIDIA in this case. Then, depending on which vendor produced the current implementation, we would run with a different number of threads, 32 for NVIDIA or 64 for AMD. After approving and merging this test, a developer from NVIDIA noticed that we had incorrectly used an \texttt{omp\_is\_initial\_device} runtime call strictly nested inside of a teams region as shown below. 

Although this test was not for the teams directive, nor for \texttt{omp\_is\_initial\_device}, this mishap led GCC to add in additional API call checks for constructs strictly nested inside teams \url{https://gcc.gnu.org/PR102972}

\begin{lstlisting}[language=C, caption=Incorrectly Strictly Nested OpenMP runtime call , label=lst:ompcpu]
#pragma omp metadirective \
     when( implementation=vendor(nvidia): \
        teams num_teams(512) thread_limit(32) ) \
    when( implementation=vendor(amd): \
        teams num_teams(512) thread_limit(64) ) \
    default (teams)
      which_device = omp_is_initial_device();
      #pragma omp distribute parallel for
         for (i = 0; i < N; i++) {
            a[i] = i;
         }
\end{lstlisting}

\subsection{Changes to the OpenMP 6.0 Specification}
\subsubsection{Discussion of Test Case Leads to Specification Issue}
In another instance, a line of questioning regarding one of our already peer reviewed and merged pull requests that came in the form of a GitHub issue led to discussion with the OpenMP Language Committee. The issue, also described here \url{https://github.com/SOLLVE/sollve_vv/issues/426} and shown in the code caption below, pointed out a unique case where a local variable is mapped to the device using a \texttt{target enter data map}, but is not explicitly mapped again on the target region itself. The stack variable is then treated as \texttt{firstprivate} in the target region and is not deallocated properly causing the stack address to be reused by a different stack variable. In this case, confusion arises due to discrepancies in the present table and produces a runtime error. The fix we agreed upon with the community member who discovered this issue is to free memory on the device associated with the stack variables before the lifetime of said variable ends on the host. Even though we were able to resolve this runtime error through deallocating the variable at the proper time, it became clear that there is no wording in the OpenMP specification that states that an OpenMP programmer must use a \texttt{target exit data} or similar directive to ensure that the lifetime of a variable does not end before it has been unmapped from a device data environment. An issue was filed with the OpenMP specification for inclusion in the 6.0 specification, but has not been resolved or merged yet.

\begin{lstlisting}[language=C, caption=Confusion surrounding lifetime of stack variable var, label=lst:ompcpu]
#pragma omp target enter data map(to: val) depend(out: val) 

#pragma omp target map(tofrom: isHost) map(alloc: h_array[0:N]) depend(inout: h_array) depend(in: val) 
  {
    isHost = omp_is_initial_device();
    for (int i = 0; i < N; ++i) {
      h_array[i] = val; // val = DEVICE_TASK1_BIT
    }
  }
\end{lstlisting}

\subsubsection{Specification Clarification from Test Case Discussion}
Further success resulted from our test case of the recently added \texttt{allocate} directive \url{https://github.com/SOLLVE/sollve_vv/pull/440}. An in-depth discussion regarding whether a certain variable could or could not be explicitly mapped, led to the inclusion of the following language in the restrictions of the threadprivate directive: ``A variable that is part of another variable (as an array element or a structure element) may appear in a threadprivate directive only if it is a static data member of a C++ class."

\subsection{Impact on Applications}
Though the SOLLVE V\&V tests primarily focus on feature tests, it also covers certain pure OpenMP application kernels extracted from applications. These tests focus on particular requirement from applications or motivated by inconsistent, regressions seen in vendor implementation, and varying support from vendors. For example, the QMCPACK application (an ECP application that is open-source, high-performance electronic structure code that implements numerous Quantum Monte Carlo (QMC) algorithms ~\cite{PKent}) found out that certain OpenMP implementations would error out when Math library functions (in \texttt{math.h}) were invoked from within a \texttt{target} region. 
In order to simplify testing over multiple implementations we created a distilled version of the same and included in our \textit{application\_kernels} directory. The developers can run these tests on the platform of their choice or refer to the SOLLVE V\&V website ~\cite{SOLLVE_web} to see if recent results for the same to verify support. 
The SOLLVE V\&V is tested across a variety of hardware platforms and OpenMP implementations regularly to document the progress of the different implementations and record any regressions.
Likewise we have looked into a number of applications like GridMINI, GESTS, LSMS etc. to collect representative application kernels which are of interest for the application developers to track. These kernels not only provide an quick and easy method for application developers to check vendor implementations, but also provide insights to vendors regarding the pain-points or critical features required by HPC applications.

%Similarly, the GridMINI application 

%\subsubsection{Potential Change from Test Case Discussion}
%A unique test we created covered a reduction on the device with two array elements from the same array. This issue is still pending on the OpenMP internal GitHub, and has generated discussion regarding whether allowing this behavior is even plausible since it is likely inefficient. See code listing below.
%
%\begin{lstlisting}[language=C, caption=Restrictions common to reduction clauses, label=lst:ompcpu]
%
 % temps[0] = 0;
%  temps[1] = 0;
%
%#pragma omp target map(tofrom: temps)
%  {
%#pragma omp parallel reduction(+:temps[0], temps[1])
%    {
%      temps[0] += 1;
%      temps[1] += 1;
 %   }
%  }
%\end{lstlisting}
  
\label{sec: communityinteraction}

\section{Related Work}
Work on OpenMP offloading has evolved in the past several years. Updated information on the various compiler tools and their coverage of OpenMP implementations especially offloading features can be found here~\cite{compilers}. 

Following are some of related works on the validation and verification of OpenMP implementations that includes features prior to offloading as well~\cite{muller2003openmp, muller2004validating,wang2012openmp,diaz2019analysis,diaz2018evaluating}. These works have been highlighting ambiguities in the specifications and reporting compiler/runtime bugs thus enabling application developers to be aware of the status of the compilers.
Another effort to test OpenMP Offloading test functions for C++ and Fortran is the OvO suite~\cite{OvO}. These tests focus on extensively testing hierarchical parallelism and mathematical functions.

Other related work includes Csmith~\cite{yang2011finding}, a comprehensive, well-cited work where the authors perform a randomized test-case generator exposing compiler bugs using differential testing. Csmith detects compiler bugs, however the strategy entails automatically mapping a randomly generated failed test to a bug that actually caused it. Such a strategy would be effective on implementations that are stable and mature. However in our case, there is frequent communication with vendors with respect to discussing and reporting bugs, and the suite also requires the use of combined and composite directives that need to be tested prior to marking a bug as a compiler or a runtime error. To that end the testsuite is not quite ready to use an approach like that used in Csmith. 

Other related work includes the parallel testsuite~\cite{dongarra1991parallel} that chooses a set of routines to test the strength of a computer system (compiler, runtime system, and hardware) in a variety of disciplines with one of the goals being to compare the ability of different Fortran compilers to automatically parallelize various loops.
The Parallel Loops test suite is modeled after the Livermore Fortran kernels~\cite{mcmahon1986livermore}. 
Overheads due to synchronization, loop scheduling and array operations are measured for the language constructs used in OpenMP in~\cite{reid2004openmp}. Significant differences between the implementations are observed, which suggested possible means of improving future performance. A microbenchmark~\cite{bull2012microbenchmark} suite was developed to measure the overhead of the task construct introduced in the OpenMP 3.0 standard, and associated task synchronization constructs. 

%The LLVM open-source compiler infrastructure~\cite{llvm} has a testing setup called \textit{lit} testing tool, which by itself does not contain accelerator (offloading) tests except for a very few tests on offloading and tasking. The LLVM testing infrastructure contains regression tests and whole programs. These regression tests are expected to always pass and should run before every commit. These tests are designed to test the various features of LLVM. Our OpenMP offloading testsuite has been loosely integrated into the LLVM lit infrastructure and we soon plan to tighten up this integration. This way the testsuite can consistently validate and verify OpenMP's offloading features being implemented within LLVM. 

These above mentioned work are some of the closely related work that focuses on tests being built and measuring overheads of implementations. There are several other related efforts that evaluate implementations using proxy, mini- or real-world applications. These work focus on mostly for performance evaluation and not the validity of the implementations. Some of these work include~\cite{davis2020performance,khalilov2021performance,pennycook2018evaluating,gayatri2018case}.

\label{sec:relatedwork}

\section{Conclusion}
Application developer teams often use OpenMP to improve the performance of their code. Newer versions of OpenMP are released every other year which include GPU offloading features, and it is vital that these features are implemented by compiler vendors \& system managers. Our testsuite ensures developers know what systems \& compilers perform the most optimally for C, C++ \& Fortran. 

The test suite has been run on multiple systems, including ORNL's Summit system and the Pre-Frontier systems Crusher with multiple compilers. Overall, it is obvious GCC, Clang \& NVHPC perform similarly for OpenMP 4.5 features, while NVHPC falls behind in later versions. LLVM's lack of a Fortran compiler makes it difficult to compare these compilers as a whole, though. On Crusher, which uses an AMD GPU, both ROCm and CCE have better support but do not progress much over version releases, especially regarding Fortran support.

Analysis of the suite's results for NVIDIA, LLVM, GNU, CCE, and ROCm on the Summit and Crusher supercomputing has shown that OpenMP 4.5 is extremely portable across the five compilers observed in this study. OpenMP 5.0 is not very portable across these compilers and some features are not supported by any of the five compilers. Developers that are concerned about portability are recommended to utilize OpenMP 4.5 features and only utilize OpenMP 5.0 features that are in the sparse set of features supported by all compilers.

Despite challenges presented to test writing, compiler implementation continues to improve over time and newer versions of OpenMP feature tests, presently supporting OpenMP 5.2, are being included in our testsuite.
\label{sec:conclusion}

\section{Acknowledgment}
This research used resources of the Oak Ridge Leadership Computing Facility at the Oak Ridge National Laboratory, which is supported by the Office of Science of the U.S. Department of Energy under Contract No. DE-AC05-00OR22725.

This research was supported by the Exascale Computing Project (17-SC-20-SC), a collaborative effort of the U.S. Department of Energy Office of Science and the National Nuclear Security Administration.
The research is also supported by the NSF under grant no. 1814609. 

%This research used resources of the National Energy Research Scientific Computing Center (NERSC), a U.S. Department of Energy Office of Science User Facility located at Lawrence Berkeley National Laboratory, operated under Contract No. DE-AC02-05CH11231.

%This manuscript has been authored by UT-Battelle, LLC under Contract No. DE-AC05-00OR22725 with the U.S. Department of Energy.  The publisher, by accepting the article for publication, acknowledges that the U.S. Government retains a non-exclusive, paid up, irrevocable, world-wide license to publish or reproduce the published form of the manuscript, or allow others to do so, for U.S. Government purposes. The DOE will provide public access to these results in accordance with the DOE Public Access Plan (http://energy.gov/downloads/doe-public-access-plan).
\label{sec:ack}

%\printbibliography
\bibliographystyle{abbrv}
\bibliography{bibliography}

\clearpage

\appendix
\section{Artifact Description}
\subsection{Abstract}
This paper explores the conformity and implementation pro- gress of various compilers for the OpenMP 4.5, 5.0 and 5.1 release specifications. Various scripts were built for testing the implementation, and to gather results from various systems. This repository includes information on the resources such as scripts,  hardware, software and other dependencies that can be used to reproduce the results that are being used in the paper.

\subsection{Artifact Availability}

Software Artifact Availability: All software is maintained in an repository under Open-Source License BSD-3.

Hardware Artifact Availability: There are no author-created hardware artifacts.

Data Artifact Availability: All data, except Crusher results are available on our website and is under Open-Source License BDS-3. Crusher results are under the discretion of OLCF.

Proprietary Artifacts: There are no author-created propritary artifacts.

List of URLs and/or DOIs where artifacts are available:

\url{https://github.com/SOLLVE/sollve\_vv}

\url{https://crpl.cis.udel.edu/ompvvsollve}

\subsection{Baseline experimental setup, and modifications made for the paper}
\subsubsection{Summit}
Relevant hardware details: [2x] IBM's 22 SIMD Multi-Core POWER9 CPUs, 512 GB of DDR4,  [6x] NVIDIA Tesla V100

Operating systems and versions: Red Hat Enterprise Linux (RHEL) version 8.2

Compilers and versions: GCC 11.2.0 IBM XL 16.1.1-10

Applications and versions: CUDA 11.5.2

Libraries and versions: OpenMP 4.5, 5.0, \& 5.1

Paper Modifications: No modifications were made

\subsubsection{Crusher}
Relevant hardware details: 64-core AMD EPYC 7A53 CPU, 512 GB of DDR4, [4x] AMD MI250X

Operating systems and versions: SUSE Linux Enterprise Server 15.3 SP3

Compilers and versions: Cray CCE 14.0.0 \& 14.0.01, AMD ROCm 4.5.0, 5.0.0, 5.1.0 \& 5.2.0

Applications and versions:  No applications were used

Libraries and versions: OpenMP 4.5, 5.0 \& 5.1

Paper Modifications: No modifications were made

\subsection{Summit results generation script}
\begin{lstlisting}[language=bash, caption= Summit script , label=lst:summit]
#!/bin/bash

#Load GCC
module load gcc/11.2.0
module cuda
module python

#run testsuite for 4.5
make CC=gcc CXX=g++ FC=gfortran LOG_ALL=1 LOG=1 VERBOSE=1 VERBOSE_TESTS=1 DEVICE_TYPE=nvidia SYSTEM=summit OMP_VERSION=4.5 all

#run testsuite for 5.0
make CC=gcc CXX=g++ FC=gfortran LOG_ALL=1 LOG=1 VERBOSE=1 VERBOSE_TESTS=1 DEVICE_TYPE=nvidia SYSTEM=summit OMP_VERSION=5.0 all

#run testsuite for 5.1
make CC=gcc CXX=g++ FC=gfortran LOG_ALL=1 LOG=1 VERBOSE=1 VERBOSE_TESTS=1 DEVICE_TYPE=nvidia SYSTEM=summit OMP_VERSION=5.1 all

#Load Clang
module use /sw/summit/modulefiles/ums/stf010/Core
module load llvm/15.0.0-20220420 #might need to change this version :)
module load cuda

#run testsuite for 4.5
make CC=clang CXX=clang++ FC=flang LOG_ALL=1 LOG=1 VERBOSE=1 VERBOSE_TESTS=1 DEVICE_TYPE=nvidia SYSTEM=summit OMP_VERSION=4.5 all

#run testsuite for 5.0
make CC=clang CXX=clang++ FC=flang LOG_ALL=1 LOG=1 VERBOSE=1 VERBOSE_TESTS=1 DEVICE_TYPE=nvidia SYSTEM=summit OMP_VERSION=5.0 all

#run testsuite for 5.1
make CC=clang CXX=clang++ FC=flang LOG_ALL=1 LOG=1 VERBOSE=1 VERBOSE_TESTS=1 DEVICE_TYPE=nvidia SYSTEM=summit OMP_VERSION=5.1 all

#Load ibm
module load xl/16.1.1-10
module load cuda

make CC=xlc CXX=xlc++ FC=xlf_r LOG_ALL=1 LOG=1 VERBOSE=1 VERBOSE_TESTS=1 DEVICE_TYPE=nvidia SYSTEM=summit OMP_VERSION=4.5 all

#run testsuite for 5.0
make CC=xlc CXX=xlc++ FC=xlf_r LOG_ALL=1 LOG=1 VERBOSE=1 VERBOSE_TESTS=1 DEVICE_TYPE=nvidia SYSTEM=summit OMP_VERSION=5.0 all

#run testsuite for 5.1
make CC=xlc CXX=xlc++ FC=xlf_r LOG_ALL=1 LOG=1 VERBOSE=1 VERBOSE_TESTS=1 DEVICE_TYPE=nvidia SYSTEM=summit OMP_VERSION=5.1 all

make report_summary
make report_json
mv report_json summit_results.json
\end{lstlisting}
\subsection{Crusher result generation commands}
\begin{lstlisting}[language=bash, caption= Crusher Commands , label=lst:crusher]
#Load rocm
ml rocm
#Load cray
ml PrgEnv-cray

#run testsuite for cce/14.0.0
ml cce/14.0.0
make CC=cc CXX=CC FC=ftn LOG=1 LOG_ALL=1 OMP_VERSION=4.5 VERBOSE=1 VERBOSE_TESTS=1 SYSTEM=crusher DEVICE_TYPE=amd all
make CC=cc CXX=CC FC=ftn LOG=1 LOG_ALL=1 OMP_VERSION=5.0 VERBOSE=1 VERBOSE_TESTS=1 SYSTEM=crusher DEVICE_TYPE=amd all
make CC=cc CXX=CC FC=ftn LOG=1 LOG_ALL=1 OMP_VERSION=5.1 VERBOSE=1 VERBOSE_TESTS=1 SYSTEM=crusher DEVICE_TYPE=amd all

#run testsuite for cce/14.0.1
ml cce/14.0.1
make CC=cc CXX=CC FC=ftn LOG=1 LOG_ALL=1 OMP_VERSION=4.5 VERBOSE=1 VERBOSE_TESTS=1 SYSTEM=crusher DEVICE_TYPE=amd all
make CC=cc CXX=CC FC=ftn LOG=1 LOG_ALL=1 OMP_VERSION=5.0 VERBOSE=1 VERBOSE_TESTS=1 SYSTEM=crusher DEVICE_TYPE=amd all
make CC=cc CXX=CC FC=ftn LOG=1 LOG_ALL=1 OMP_VERSION=5.1 VERBOSE=1 VERBOSE_TESTS=1 SYSTEM=crusher DEVICE_TYPE=amd all

#run testsuite for rocm/4.5.0
module load PrgEnv-amd
module load rocm/4.5.0
make CC=amdclang CXX=amdclang++ FC=ftn LOG=1 LOG_ALL=1 OMP_VERSION=4.5 VERBOSE=1 VERBOSE_TESTS=1 SYSTEM=crusher DEVICE_TYPE=amd all
make CC=amdclang CXX=amdclang++ FC=ftn LOG=1 LOG_ALL=1 OMP_VERSION=5.0 VERBOSE=1 VERBOSE_TESTS=1 SYSTEM=crusher DEVICE_TYPE=amd all
make CC=amdclang CXX=amdclang++ FC=ftn LOG=1 LOG_ALL=1 OMP_VERSION=5.1 VERBOSE=1 VERBOSE_TESTS=1 SYSTEM=crusher DEVICE_TYPE=amd all

#run testsuite for rocm/5.0.0
module load rocm/5.0.0
make CC=amdclang CXX=amdclang++ FC=ftn LOG=1 LOG_ALL=1 OMP_VERSION=4.5 VERBOSE=1 VERBOSE_TESTS=1 SYSTEM=crusher DEVICE_TYPE=amd all
make CC=amdclang CXX=amdclang++ FC=ftn LOG=1 LOG_ALL=1 OMP_VERSION=5.0 VERBOSE=1 VERBOSE_TESTS=1 SYSTEM=crusher DEVICE_TYPE=amd all
make CC=amdclang CXX=amdclang++ FC=ftn LOG=1 LOG_ALL=1 OMP_VERSION=5.1 VERBOSE=1 VERBOSE_TESTS=1 SYSTEM=crusher DEVICE_TYPE=amd all

#run testsuite for rocm/5.1.0
module load rocm/5.1.0
make CC=amdclang CXX=amdclang++ FC=ftn LOG=1 LOG_ALL=1 OMP_VERSION=4.5 VERBOSE=1 VERBOSE_TESTS=1 SYSTEM=crusher DEVICE_TYPE=amd all
make CC=amdclang CXX=amdclang++ FC=ftn LOG=1 LOG_ALL=1 OMP_VERSION=5.0 VERBOSE=1 VERBOSE_TESTS=1 SYSTEM=crusher DEVICE_TYPE=amd all
make CC=amdclang CXX=amdclang++ FC=ftn LOG=1 LOG_ALL=1 OMP_VERSION=5.1 VERBOSE=1 VERBOSE_TESTS=1 SYSTEM=crusher DEVICE_TYPE=amd all

#run testsuite for rocm/5.2.0
module load rocm/5.2.0
make CC=amdclang CXX=amdclang++ FC=ftn LOG=1 LOG_ALL=1 OMP_VERSION=4.5 VERBOSE=1 VERBOSE_TESTS=1 SYSTEM=crusher DEVICE_TYPE=amd all
make CC=amdclang CXX=amdclang++ FC=ftn LOG=1 LOG_ALL=1 OMP_VERSION=5.0 VERBOSE=1 VERBOSE_TESTS=1 SYSTEM=crusher DEVICE_TYPE=amd all
make CC=amdclang CXX=amdclang++ FC=ftn LOG=1 LOG_ALL=1 OMP_VERSION=5.1 VERBOSE=1 VERBOSE_TESTS=1 SYSTEM=crusher DEVICE_TYPE=amd all

\end{lstlisting}
\subsection{Sample Result Output}
Shown below is a single test result sampled from the results json file generated by the Crusher results generation commands 
\begin{lstlisting}[caption= Sample results json output , label=lst:sample_rsult]
{
"Binary path": "bin/alpaka_complex_template.cpp",
"Compiler command": "amdclang++ -I./ompvv -std=c++11 -lm -O3 -fopenmp -fopenmp -fopenmp-targets=amdgcn-amd-amdhsa -Xopenmp-target=amdgcn-amd-amdhsa -march=gfx90a  -D__NO_MATH_INLINES -U__SSE2_MATH__ -U__SSE_MATH__",
"Compiler ending date": "Thu 14 Jul 2022 04:30:15 PM EDT",
"Compiler name": "amdclang++ AMD clang version 13.0.0 (https://github.com/RadeonOpenCompute/llvm-project roc-4.5.0 21422 e2489b0d7ede612d6586c61728db321047833ed8)",
"Compiler output": "",
"Compiler result": "PASS",
"Compiler starting date": "Thu 14 Jul 2022 04:30:03 PM EDT",
"OMP version": "4.5",
"Runtime ending date": "Thu 14 Jul 2022 04:30:15 PM EDT",
"Runtime only": false,
"Runtime output": "\u001b[0;32m \n\n running: bin/alpaka_complex_template.cpp.run \u001b[0m\nalpaka_complex_template.cpp.o: PASS. exit code: 0\n\u001b[0;31malpaka_complex_template.cpp.o:\n[OMPVV_INFO: alpaka_complex_template.cpp:40] Test is running on device.\n[OMPVV_INFO: alpaka_complex_template.cpp:58] The value of errors is 0.\n[OMPVV_RESULT: alpaka_complex_template.cpp] Test passed on the device.\u001b[0m\n",
"Runtime result": "PASS",
"Runtime starting date": "Thu 14 Jul 2022 04:30:14 PM EDT",
"Test comments": "none",
"Test gitCommit": "98cae2b",
"Test name": "alpaka_complex_template.cpp",
"Test path": "tests/4.5/application_kernels/alpaka_complex_template.cpp",
"Test system": "crusher"
}
\end{lstlisting}

\end{document}